\documentclass[twocolumn, twocolappendix]{aastex631}
\usepackage{float}
\usepackage{wasysym}
\usepackage{booktabs}
\usepackage{changepage}
\usepackage{xspace}
\usepackage[encapsulated]{CJK}
\usepackage{savesym}
\savesymbol{iint}
\savesymbol{iiint}
\usepackage{amsmath}	
\restoresymbol{ams}{iint}
\restoresymbol{ams}{iiint}

\defcitealias{Weibel2026}{W26} 

\begin{document}


\begin{abstract}

Observational campaigns with JWST have revealed a higher-than-expected abundance of UV-bright galaxies at $z\gtrsim10$, with various proposed theoretical explanations. A powerful complementary constraint to break degeneracies between different models is galaxy clustering. In this paper, we combine PANORAMIC pure parallel and legacy imaging along 34 independent sightlines to measure the cosmic variance ($\sigma_{\rm CV}$) in the number counts of Lyman break galaxies at $z\sim10$ which is directly related to their clustering strength. We find $\sigma_{\rm CV}=0.96^{+0.20}_{-0.18}$, $1.46^{+0.54}_{-0.44}$, and $1.71^{+0.72}_{-0.59}$ per NIRCam pointing ($\sim9.7\,{\rm arcmin}^2$, $\lesssim1.5\,{\rm pMpc}$ at $z\sim10$) for galaxies with M$_{\rm UV}<-19.5$, $-20$, and $-20.5$. Comparing to galaxies in the fiducial \texttt{UniverseMachine}, we find that $\sigma_{\rm CV}$ is consistent with our measurements, but that the number densities are a factor $\gtrsim5$ lower. We implement simple models in the \texttt{UniverseMachine} that represent different physical mechanisms to enhance the number density of UV-bright galaxies. All models decrease $\sigma_{\rm CV}$ by placing galaxies at fixed M$_{\rm UV}$ in lower mass halos, but to varying degrees. Combined constraints on $\sigma_{\rm CV}$ and the UVLF thus tentatively disfavor models that globally increase the star formation efficiency (SFE) or the scatter in the M$_{\rm UV}$-$M_{\rm halo}$ relation, while models that decrease the mass-to-light ratio, or assume a power-law scaling of the SFE with $M_{\rm halo}$ agree better with the data. We show that with sufficient additional independent sightlines, robust discrimination between models is possible, paving the way for powerful constraints on the physics of early galaxy evolution through NIRCam pure parallel imaging.

\end{abstract}


\title{Exploring Cosmic Dawn with PANORAMIC II: Cosmic Variance and Galaxy Clustering at $\mathbf{z\sim10}$}

\author[0000-0001-8928-4465]{Andrea Weibel}
\altaffiliation{These authors contributed equally to this work.}
\affiliation{Department of Astronomy, University of Geneva, Chemin Pegasi 51, 1290 Versoix, Switzerland}

\author[0000-0002-8896-6496, gname=Christian Kragh, sname=Jespersen]{Christian Kragh Jespersen}
\altaffiliation{These authors contributed equally to this work.}
\affiliation{Department of Astrophysical Sciences, Princeton University, Princeton, NJ 08544, USA}

\author[0000-0001-5851-6649]{Pascal A.\ Oesch}
\affiliation{Department of Astronomy, University of Geneva, Chemin Pegasi 51, 1290 Versoix, Switzerland}
\affiliation{Cosmic Dawn Center (DAWN), Denmark}
\affiliation{Niels Bohr Institute, University of Copenhagen, Jagtvej 128, K{\o}benhavn N, DK-2200, Denmark}

\author[0000-0003-2919-7495]{Christina C.\ Williams}
\affiliation{NSF NOIRLab, 950 N. Cherry Ave., Tucson, AZ 85719, USA}
\affiliation{Steward Observatory, University of Arizona, 933 North Cherry Avenue, Tucson, AZ 85721, USA}

\author[0000-0001-5063-8254]{Rachel Bezanson}
\affiliation{Department of Physics and Astronomy and PITT PACC, University of Pittsburgh, Pittsburgh, PA 15260, USA}

\author[0000-0003-2680-005X]{Gabriel Brammer}
\affiliation{Cosmic Dawn Center (DAWN), Denmark}
\affiliation{Niels Bohr Institute, University of Copenhagen, Jagtvej 128,
K{\o}benhavn N, DK-2200, Denmark}

\author[0000-0001-9978-2601]{Aidan P. Cloonan}
\affiliation{Department of Astronomy, University of Massachusetts, Amherst, MA 01003, USA}

\author[0000-0001-8460-1564]{Pratika Dayal}
\affiliation{Canadian Institute for Theoretical Astrophysics, 60 St George St, University of Toronto, Toronto, ON M5S 3H8, Canada}
\affiliation{David A. Dunlap Department of Astronomy and Astrophysics, University of Toronto, 50 St George St, Toronto ON M5S 3H4, Canada}
\affiliation{Department of Physics, 60 St George St, University of Toronto, Toronto, ON M5S 3H8, Canada}

\author[0000-0003-3760-461X]{Anne Hutter}
\affiliation{Institute for Astronomy, University of Vienna, Türkenschanzstrasse 17, A-1180 Vienna, Austria}

\author[0000-0001-7673-2257]{Zhiyuan Ji}
\affiliation{Steward Observatory, University of Arizona, 933 North Cherry Avenue, Tucson, AZ 85721, USA}

\author[0000-0003-0695-4414]{Michael V. Maseda}
\affiliation{Department of Astronomy, University of Wisconsin-Madison, 475 N. Charter St., Madison, WI 53706 USA}

\author[0000-0002-7087-0701]{Marko Shuntov}
\affiliation{Cosmic Dawn Center (DAWN), Denmark}
\affiliation{Niels Bohr Institute, University of Copenhagen, Jagtvej 128, K{\o}benhavn N, DK-2200, Denmark}
\affiliation{Department of Astronomy, University of Geneva, Chemin Pegasi 51, 1290 Versoix, Switzerland}

\author[0000-0001-7160-3632]{Katherine E. Whitaker}
\affiliation{Department of Astronomy, University of Massachusetts, Amherst, MA 01003, USA}
\affiliation{Cosmic Dawn Center (DAWN), Denmark}

\correspondingauthor{Andrea~Weibel\,(andrea.weibel@unige.ch)~\&\\~Christian~Kragh~Jespersen\,(ckragh@princeton.edu)}

\section{Introduction}
\label{sec:intro}

One of the biggest surprises that the James Webb Space Telescope (JWST) has brought us over the last few years is the high number of luminous galaxies at redshifts $z\gtrsim10$ that it revealed through deep imaging \citep[e.g.][]{Naidu2022,Castellano2022,Adams2023,Finkelstein2023,Atek2023,Casey2024,Hainline2024}, and confirmed spectroscopically  \citep[e.g.][]{ArrabalHaro2023,Curtis-Lake2023,Castellano2024,Fujimoto2024,Harikane2024,Carniani2024,Naidu2026}. Statistical measurements of the UV luminosity function (UVLF) at $z\gtrsim10$ from various imaging programs and fields \citep[e.g.][]{Donnan2023,Perez-Gonzalez2023b,Harikane2023,Adams2024,Finkelstein2024,Robertson2024,Donnan2024,Whitler2025,Weibel2026} therefore yielded an excess at the bright end, and a shallow evolution with redshift compared to (extrapolations of) pre-JWST measurements \citep[e.g.][]{Bouwens2016, Ishigaki2018, Oesch2018}, and theoretical expectations \citep[e.g.][]{Dayal2014, Mason2015, Tacchella2018, Williams2018, Behroozi2019, Steinhardt2021}.

A wealth of models that can explain this high abundance of UV-bright galaxies have been discussed in the literature. This includes models that boost the observed UV-luminosity of galaxies at fixed halo mass through e.g. a more top-heavy initial mass function \citep[IMF;][]{Yung2024, Trinca2024, Cueto2024, Lu2025, Mauerhofer2025, Hutter2025}, a higher star formation efficiency \citep[SFE;][]{Dekel2023,Li2024,Boylan-Kolchin2025,Mauerhofer2025,Somerville2025}, a lack of dust attenuation \citep{Ferrara2023}, or a contribution of active galactic nuclei (AGN) to the UV-luminosity \citep{Pacucci2022,Hegde2024}. Using the FIREBox$^{HR}$ simulation, \citet{Feldmann2025} showed that the UVLF and the UV luminosity density at $z\sim6-14$ can also be reproduced based on just a weakly halo mass dependent SFE. A different class of models increases the scatter in the relation between the absolute UV-magnitude (M$_{\rm UV}$) and the halo mass ($M_{\rm halo}$), in particular through enhanced burstiness in the star formation histories (SFHs) of high redshift galaxies \citep{Mason2023,Shen2023,Sun2023,Kravtsov2024}. Crucially, while these models all boost the number density of galaxies, they have different implications for their clustering strength. Hence they can in principle be distinguished based on clustering measurements \citep[e.g.][]{Munoz2023,Gelli2024,Shuntov2025b}.

Observational results regarding the abundance of UV-bright galaxies at $z\gtrsim10$ are usually based on a small number of independent lines of sight. They are therefore subject to significant cosmic variance, which describes fluctuations in the number density of galaxies because of their clustering in the cosmic web \citep[e.g.][]{Moster2011,Jespersen2025}. Indeed, determinations of the UVLF based on a larger number of independent sightlines, and/or using data outside of the commonly studied legacy fields found somewhat lower number densities at $z\gtrsim10$ \citep[e.g.][]{Willott2024,Morishita2025,Asada2026}. Nevertheless, the impact of cosmic variance on JWST-based results remains large due to its limited field of view, and the fact that many studies build on a small number of legacy fields containing most of the data.

\citet[][\citetalias{Weibel2026} hereafter]{Weibel2026} recently combined legacy imaging with pure parallel imaging from PANORAMIC (GO-2514, PIs Williams \& Oesch, \citealt{Williams2025}) to measure UVLFs at $z\sim10$, $z\sim13$, and $z\sim17$ along 35 independent lines of sight. This work confirmed the high abundance of UV-bright (M$_{\rm UV}\lesssim-21$) galaxies at $z\sim10$ and $z\sim13$, but found a more rapid evolution in the UVLF than previous studies, corresponding to a drop in the UV luminosity density by a factor $>50\times$ from $z\sim10$ to $z\sim17$. Many recently proposed theoretical models are consistent with these findings but cannot be distinguished solely based on the UVLF.

One way to break degeneracies between models is to measure the clustering strength of galaxies, so as to relate them to the dark matter halos they reside in. Observationally, galaxy clustering is usually quantified via the (angular) two-point correlation function (2PCF). This has been measured in the local Universe and at low redshifts from large ground-based surveys \citep[e.g.][]{Norberg2002, Budavari2003, Zehavi2005, Li2006}, and later out to $z\sim1$ \citep[e.g.][]{Meneux2009}, $z\sim2$ \citep[e.g.][]{Blanc2008, Wake2011, Lin2012, McCracken2015}, and $z\sim3-8$ \citep[e.g.][]{Hildebrandt2009,Barone-Nugent2014, Harikane2016, Ishikawa2017, Harikane2018,Ye2025}. Recent works based on JWST imaging data have extended measurements of the 2PCF out to  $z\sim11$ \citep{Dalmasso2024} and $z\sim12$ \citep{Paquereau2025} with photometric redshifts informed by deep NIRCam imaging, or made use of NIRCam/grism surveys providing spectroscopic redshifts out to $z\sim8$ \citep{Shuntov2025b}. 

On the other hand, the clustering behavior of dark matter halos is relatively well understood based on cosmological N-body simulations \citep[e.g.][]{Springel2005}, and analytical models \citep[e.g.][]{Cooray2002}. Combining this with a so-called halo occupation distribution \citep[HOD; e.g.][]{Berlind2002}, one can then relate halos to galaxies, and model the galaxy 2PCF. Inversely, the HOD can be inferred from a measurement of the 2PCF \citep[see e.g.][]{Paquereau2025}.

A quantity that is used to characterize the clustering of halos is the so-called halo bias which is defined as the excess clustering of halos relative to the matter density field. Analogously, the galaxy bias is defined as the excess clustering of galaxies relative to the matter density field as a function of the physical scale $r$. On sufficiently large scales, the halo bias and thus the galaxy bias is scale-independent, and only a function of halo mass \citep{Mo1996}. This bias is referred to as the linear bias because it is valid in the linear regime, where fluctuations in the matter field can be approximated as linear deviations from the mean density of the Universe.

Clustering of halos and galaxies is stronger on smaller scales, also referred to as the non-linear regime, where the (non-linear) physics of the collapse of halos into virialized structures, and of galaxy formation become important. Various works show that non-linear effects still enhance the clustering of halos on relatively large, so-called quasi-linear scales (see e.g. \citealt{Jose2016} and references therein). Building on their results for halos, \citet{Jose2017} investigated the clustering of Lyman break galaxies (LBGs) at $z=3-5$ and found that on angular scales of $\sim5-100$\arcsec\ their clustering is enhanced by an order of magnitude due to non-linear effects. 

\citet{Robertson2010} suggested that the galaxy bias of high redshift LBGs could be constrained from measurements of cosmic variance, designated as $\sigma_{\rm CV}$ hereafter. They define a galaxy bias as

\begin{equation}
\label{eq:bias_cv}
b_{g,\,\rm CV} = \frac{\sigma_{\rm CV}}{\sigma_{\rm DM}}
\end{equation}

where $\sigma_{\rm DM}$ is the field-to-field variance in the underlying dark matter density field. Importantly, $\sigma_{\rm CV}$ and thus $b_{g,\,\rm CV}$ incorporates clustering on all scales smaller than the probed survey volume. This means that it significantly differs from the linear bias as, e.g., measured from the 2PCF, because it includes stronger clustering on non-linear scales. Equation \ref{eq:bias_cv} has been successfully applied in the literature to measure the clustering of galaxies of different types out to $z\sim2$ \citep{Lopez-Sanjuan2015, Cameron2019}.

In this work, we use the sample of LBGs compiled in \citetalias{Weibel2026} to measure $\sigma_{\rm CV}$ for UV-bright galaxies at $z\sim10$. This can be related to the clustering strength of these galaxies through Equation \ref{eq:bias_cv}, which can in turn break degeneracies between different models that have been proposed to explain the high abundance of UV-bright galaxies at $z\sim10$. The paper is structured as follows:

We present the data as well as the methods applied to measure cosmic variance in Section \ref{sec:methods}. Our results are presented in Section \ref{sec:results}, starting with the measurements of $\sigma_{\rm CV}$. We then describe our implementation of various simple models that boost the number density of UV-bright galaxies at $z\sim10$ in the \texttt{UniverseMachine} \citep{Behroozi2019}, and investigate their impact on $\sigma_{\rm CV}$. This is followed by a discussion of our results in Section \ref{sec:discussion} which includes a quantification of the effect of non-linear scales on the measured $\sigma_{\rm CV}$, and future prospects for $\sigma_{\rm CV}$ constraints based on additional pure parallel imaging. We summarize our findings and conclude in Section \ref{sec:summary_conclusions}.

Throughout this work we assume a $\Lambda$CDM cosmology with parameters from the nine-year Wilkinson Microwave Anisotropy Probe Observations \citep{WMAP9}, $h=0.6932$ and $\Omega_{m,0}=0.2865$. All magnitudes are reported in the AB system \citep{Oke1983}. 

\section{Data and Methods}
\label{sec:methods}

The data used in this work are largely identical to the data described in \citetalias{Weibel2026}. We provide a brief summary below, and refer the reader to \citetalias{Weibel2026} for details.

\subsection{Imaging and Photometry}

This work builds on JWST and HST imaging mosaics from various legacy surveys, retrieved from the DAWN JWST Archive (DJA\footnote{\url{https://dawn-cph.github.io/dja/imaging/v7/}}), as well as pure parallel imaging from the first PANORAMIC data release\footnote{\url{https://panoramic-jwst.github.io/}} \citep{Williams2025}. PANORAMIC is a JWST cycle 1 pure parallel NIRCam imaging survey. The pure parallel observing mode enables the use of NIRCam while a different instrument is used to carry out primary observations, implying pointing locations at a fixed offset from the respective primary pointing. While this yields some imaging that overlaps with or is in close proximity to extragalactic legacy fields, it crucially adds independent pointings that follow a nearly random distribution on the sky due to physically unrelated primary targets. Specifically, PANORAMIC provides 28 such independent sightlines, despite initial technical difficulties (see \citealt{Williams2025} for more details). All mosaics have been reduced with \texttt{grizli} \citep{10.5281/zenodo.8370018}, as described in e.g., \citet{Valentino2023}. A list of all JWST programs contributing imaging data to the legacy fields is provided in Section 2.1 of \citetalias{Weibel2026} and the PANORAMIC mosaics are available as high-level science products on MAST\footnote{\dataset[doi:10.17909/fpzr-as35
]{https://doi.org/10.17909/fpzr-as35}}.

Photometric catalogs are generated following the methods outlined in \citet{Weibel2024} and \citet{Williams2025}. We detect sources in an inverse-variance weighted stack of the NIRCam/LW filters F277W, F356W, and F444W using \texttt{SourceExtractor} \citep{Bertin1996}. Running \texttt{SourceExtractor} in dual mode, we measure fluxes on point spread function (PSF) matched versions of the mosaics in each filter through circular apertures with a radius of 0.16\arcsec, and scale these fluxes to total based on the Kron aperture in the detection image, and an additional correction to account for flux in the wings of the PSF \citep[see e.g.][]{Skelton2014, Whitaker2019, Weaver2024}.

\subsection{Galaxy Sample}
\label{sec:sample}

\citetalias{Weibel2026} selected LBGs as F115W, F150W, and F200W dropouts corresponding to $z\sim10$, $z\sim13$, and $z\sim17$. We only use the F115W dropout sample here which consists of 86 galaxies at $z\sim10$. The sample selection mostly relies on conservative color cuts defined as 

\begin{equation}
\begin{gathered}
\label{eq:dropout_sel}
    {\rm F115W} - {\rm F150W} > 1.5\,\,\,\,\land\\[1ex]
    {\rm F150W} - {\rm F356W} < 0.5\,\,\,\,\land\\[1ex]
    {\rm SNR(F150W)} > 8\,\,\,\,\land\\[1ex]
    {\rm SNR(F200W)} > 3
\end{gathered}
\end{equation}

where FXXXW is the AB-magnitude in each filter, and we apply a 2$\sigma$ upper limit to the flux in F115W before computing the first color. This selection is complemented with the \texttt{use\_phot} flag from \citet{Weibel2024}. We further reject confident low redshift interlopers based on \texttt{eazy} \citep{Brammer2008}, and perform a visual inspection to remove remaining diffraction spikes, spurious detections, and sources with clear detected flux below the supposed Lyman break. We exclude data from the Abell-2744 cluster field in this work to avoid complications introduced to flux limits and volumes by gravitational lensing, which reduces the sample size to 74 galaxies.

Of these 74 galaxies, 24 have publicly available spectroscopic redshifts on the DJA. Reassuringly, all except for one of them are confirmed to be at $z>8.6$. As discussed in Section 2.5 of \citetalias{Weibel2026}, the only exception may be a high redshift galaxy that is blended with a lower redshift ($z=1.98$) foreground galaxy \citep[see][]{Bunker2024}. Further, while spectroscopic redshifts are preferentially available for sources in legacy fields with deep photometry in many bands, the success rate of at least $\sim96$\%\ for sources with $z_{\rm spec}$ nevertheless suggests a high purity of the sample.

All galaxies are fit with \texttt{bagpipes} \citep{Carnall2018}, with a uniform redshift prior across the nominal range of the dropout selection plus (minus) 0.1 at the upper (lower) end, $z\in(8.5, 11.4)$. To consistently infer the redshift, UV-magnitude M$_{\rm UV}$, and -slope $\beta$ for each source within a Bayesian framework, we sample 500 spectral energy distributions (SEDs) from the \texttt{bagpipes} posterior, and measure $z$, M$_{\rm UV}$, and $\beta$ from each to get our final estimates and uncertainties of these quantities as the median, 16th and 84th percentiles.

\subsection{Completeness Correction}
\label{sec:compl_corr}

Again following \citetalias{Weibel2026}, we determine the completeness of the sample in two steps. First, we assess the detection completeness as a function of the source magnitude in the detection image through injection-recovery simulations with the GaLAxy survey Completeness AlgoRithm 2 (\texttt{GLACiAR2}) software \citep{Carrasco2018, Leethochawalit2022}. Second, we estimate the selection completeness by creating mock SEDs based on the redshift and UV-properties from the 500 SEDs sampled from the \texttt{bagpipes} posterior. For each of the 500 sets of $z$, M$_{\rm UV}$, and $\beta$, we produce a mock SED with a power-law shape, and create 1000 realizations of the photometry by adding Gaussian scatter to the synthetic flux in each band according to the flux uncertainties specified in the photometric catalog. The selection completeness is then given by the number of the 1000 realizations that would pass the selection cut in Equation \ref{eq:dropout_sel}, divided by the number that would be detected by \texttt{SourceExtractor}, according to the combined SNR in F277W, F356W, and F444W. In this way, the two completeness values are independent and can be multiplied to obtain a total completeness for each source. We further convert the values in the weight maps from the DJA to an M$_{\rm UV}$ limit per pixel following \citetalias{Weibel2026}, and derive a median M$_{\rm UV}$ limit per field.

\subsection{Cosmic Variance Measurements}
\label{sec:cosmic_variance_measurement}

We follow two different approaches to empirically constrain cosmic variance at $z\sim10$: bootstrapping and Bayesian forward modeling. We start from PANORAMIC pointings outside of any legacy fields with $>2\,$deg separation on the sky, so that they can be considered uncorrelated in terms of galaxy clustering\footnote{We choose this somewhat arbitrary scale to be roughly twice the angular BAO scale at $z\sim10$. At these scales, any bias incurred from large-scale density modes will be at the percent-level, much below our sensitivity.}. There are 28 such pointings, most of which consist of a single NIRCam field of view (see \citealt{Williams2025}). To include the legacy fields as additional lines of sight without accounting for the more complex survey geometries and substantially larger area per field, we create ``mock'' NIRCam pointings by placing the NIRCam field of view at a random position and with a random position angle on top of each legacy field mosaic. We assert that at least 90\% of the footprint are covered by data in all required filters (see Section \ref{sec:sample}) and then treat the sampled part of the legacy field as an independent field. We proceed in this way for the UDS, COSMOS, EGS, and the two GOODS fields (North and South). The NEP-TDF consists of a single NIRCam pointing and can therefore be included as is. This yields 34 independent lines of sight in total, with roughly identical survey geometries, i.e. single NIRCam pointings of $\sim9.7$arcmin$^2$.

Next, we wish to quantify the cosmic variance for different M$_{\rm UV}$ limits. For a given limit, we include only brighter galaxies and restrict the analysis to fields with sufficient median depth. Consequently, the number of contributing fields decreases toward fainter limits, while the number of galaxies decreases toward brighter limits. Given these constraints, we adopt M$_{\rm UV}<-19.5$, $-20$, and $-20.5$ as feasible limits, corresponding to 9, 17, and 29 independent fields, and 57, 37, and 19 galaxies respectively. To account for remaining small differences in the survey area of different fields, we scale the observed number of galaxies in each field to the nominal NIRCam field of view of 9.7 arcmin$^2$.

\subsubsection{Bootstrapping}
\label{sec:bootstrapping}

The field-to-field variance in the number counts comes from the combined effect of Poisson noise and cosmic variance. To separate out the latter, we use the relation

\begin{equation}
\label{eq:cv_bootstrap}
\sigma_{\rm CV}^2 = \frac{\sigma^2_{\mathrm{total}}-\sigma^2_{\mathrm{Poisson}}}{\mu^2} = \frac{\langle N^2\rangle - \langle N\rangle^2 - \langle N \rangle} {\langle N \rangle^2}
\end{equation}
 
following e.g. \citet{Somerville2004} and \citet{Moster2011} where N stands for the completeness corrected galaxy number counts per field, $\mu=\langle N\rangle$ is the mean number count, and we make use of the fact that the Poisson contribution to the variance is $\sigma_{\rm Poisson}^2={\rm \langle N\rangle}$. Note that $\sigma_{\rm CV}$ is defined as a fractional variance.

Equation \ref{eq:cv_bootstrap} can be directly applied to the numbers of galaxies per field for the three M$_{\rm UV}$ limits respectively. To estimate uncertainties for the inferred cosmic variance, we perform 1000 bootstrapped measurements of $\sigma_{\rm CV}$ by sampling 34 of the 34 fields each time, allowing for fields to be sampled multiple times. To additionally leverage the larger survey area covered by the legacy fields (CEERS, COSMOS, UDS, GOODS-N, and GOODS-S), we sample a new random NIRCam pointing within each field's footprint in each iteration. We then take the median as well as the 16th and 84th percentile of the 1000 measurements as our final bootstrapped values and uncertainties for $\sigma_{\rm CV}$.

\subsubsection{MCMC-Fitting}
\label{sec:mcmc_fitting}

A more principled approach for quantifying the cosmic variance is to directly fit to the galaxy number count distribution. To this end, we model the observed distribution of completeness-corrected galaxy counts across independent fields and for our different M$_{\rm UV}$ limits using a Bayesian Markov Chain Monte Carlo (MCMC) framework. We assume that the underlying field-to-field distribution of galaxy counts follows a Gamma distribution \citep{Steinhardt2021, Jespersen2025}. The Gamma distribution is characterized by a mean and a variance, which, in contrast to the Poisson distribution, can be specified independently of each other, allowing for a super-Poissonian variance. In the limit that the variance becomes equal to the mean, the Gamma distribution closely resembles the Poisson distribution. Therefore, the Gamma distribution, with variance equal to $\sigma_{\mathrm{Poisson}}^2+\mu^2\sigma_{\mathrm{CV}}^2$ goes towards the expected Poisson distribution in the limit of no cosmic variance/clustering, with the $\sigma_{\mathrm{CV}}$-term encoding exactly the cosmic variance/clustering we wish to constrain.\footnote{Another way to think of the Gamma distribution is as the distribution arising from the theoretical expectation for positive-definite variables (galaxy counts) with stochastic large-scale structure modulation.}. Other distributions, such as the negative binomial distribution, are also possible adequate choices since they also share the same characteristic --- Poisson-like in the absence of cosmic variance, but allowing for the inclusion of cosmic variance. However, the difference between adopting either the Gamma or Negative Binomial is empirically negligible \citep{Jespersen2025}.

We infer the mean and variance parameters\footnote{In practice, the input parameters to the Gamma distribution are a shape and a scale parameter, but these are directly related to the mean and variance, making their use in inference equivalent.} of the Gamma distribution via MCMC sampling using the \texttt{emcee} sampler \citep{emcee}. Critically, the likelihood function we adopt down-weights variance contributions from rare, highly over-dense fields. This is achieved by defining the likelihood as the Negative Binomial distribution across the binned number counts, which in essence acts as an outlier-resistant Poisson-likelihood, as per the recommendation (and typical best-practice approach) of \cite{Hogg2010_fittingmodels}. This approach reduces the influence of outlier fields, in line with the findings of \cite{Jespersen2025}, who find that empirically calibrated cosmic variances are biased towards high values. This is due to the highly skewed nature of the distribution of galaxy number counts across different fields.\footnote{To clarify, the skewness implies that the galaxy number count distribution is highly asymmetric, having a preference for both very small and very large values, depopulating the central region that usually stabilizes fits.} We adopt the maximally non-informative prior distribution for the variance ($p(\sigma) \propto \sigma^{-1}$, also known as the \textit{Jeffrey's prior}), which effectively forces the variance to be small unless a high variance is strongly favored by the data \citep{jeffreys1946invariant}. We also adopt a weak Gaussian prior for the mean, centered on the sample mean and with a width of 30\% of the sample mean. This mainly helps guide the sampler and does not have any meaningful effect on our inference results.

In addition to the real data, we apply the same measurement to galaxies in the \texttt{UniverseMachine} \citep{Behroozi2019}, a parametric empirical model built with a set of differential equations governing galaxy evolution on pre-computed merger trees from the Bolshoi-Planck dark matter simulation \citep{Klypin2016,Rodriguez-Puebla2016}. It provides eight independent mock light-cones for the survey geometries of the five CANDELS fields \citep{Grogin2011, Koekemoer2011}. We use four out of the five, since one has unexplainable and significantly lower typical number densities, giving us access to 32 independent ``lines of sight'' of area typically a bit below a square degree. Further, the \texttt{UniverseMachine} was directly calibrated on observational constraints, including cross-correlation functions at $0\leq z\leq0.13$, auto-correlation functions at $z\leq0.7$, stellar mass functions out to $z\sim8$, as well as cosmic star formation rate densities and UVLFs out to $z\sim10$.

We measure the mean number of galaxies and their variance for the different M$_{\rm UV}$ limits within light cones in the \texttt{UniverseMachine} whose size is given by the size of a NIRCam pointing on the sky and extends over $9.2<z<10.9$. This is slightly smaller than the nominal redshift range of the F115W dropout bin ($8.6<z<11.3$, see \citetalias{Weibel2026}) to approximate the redshift selection function which declines towards the edges of the nominal range. We emphasize that the resulting cosmic variance includes contributions from clustering on all scales smaller than the survey volume, including non-linear scales. Thanks to the consistent measurement method, we can however directly compare the values inferred from the \texttt{UniverseMachine} to those inferred from the data. We integrate the dark matter power spectrum over the same survey volume, to estimate the variance in the underlying dark matter distribution using a python version of the code \texttt{QUICKCV} \citep{Newman2014}, described in \citet{Newman2002}, to obtain $\sigma_{\rm DM}=0.031$. With this, we can turn our estimated values of $\sigma_{\rm CV}$ into estimates of the (cosmic variance based) galaxy bias $b_{g,\,\rm CV}$ using Equation \ref{eq:bias_cv}. We note that a slight increase or decrease in the assumed redshift range that is used to approximate the LBG selection function does not have a critical impact on $\sigma_{\rm DM}$. For example, assuming the maximal nominal range, $8.6<z<11.3$ yields $\sigma_{\rm DM,\,min}=0.025$, while a narrower range of $9.3<z<10.3$ yields $\sigma_{\rm DM,\,max}=0.039$, reflecting changes of $<30$\%.

\section{Results}
\label{sec:results}

\subsection{Observational Constraints}
\label{sec:cv_results}

We start by presenting our observational constraints on cosmic variance, and the implied galaxy bias, both from bootstrapping Equation \ref{eq:cv_bootstrap}, and the MCMC-fitting approach.

\subsubsection{Bootstrapping}
\label{sec:bootstrapping_results}

From our 1000 bootstrapped samples we find the following values for the three M$_{\rm UV}$ limits: $\sigma_{\rm CV,\,bootstr.}=0.96_{-0.26}^{+0.28}$, $1.76_{-0.43}^{+0.50}$, and $2.37_{-0.44}^{+0.64}$. As discussed in Section \ref{sec:mcmc_fitting}, the underlying number count distributions are highly skewed, and truncated at 0 which implies that the measured variance is likely biased high. This effect is more significant for small number counts,  and if the intrinsic (cosmic) variance is higher, as is expected for more massive galaxies. In other words, if UV-bright galaxies at $z\sim10$ are strongly clustered, the expected intrinsic distribution of number counts per NIRCam pointing will have a long tail out to high number counts. Any measurement of the variance of that distribution based on small number statistics is therefore expected to be biased high due to a small number of outliers boosting the variance (see \citealt{Jespersen2025}). A test on how much this affects the cosmic variance values inferred through bootstrapping is provided by our second approach to measuring $\sigma_{\rm CV}$.

\subsubsection{MCMC Fitting}
\label{sec:mcmc_results}

MCMC fitting of a Gamma distribution to the measured number counts per field yields $\sigma_{\rm CV}= 0.96^{+0.20}_{-0.18}$, $1.46^{+0.54}_{-0.44}$, $1.71^{+0.72}_{-0.59}$, for M$_{\rm UV}<-19.5$, $-20$, and $-20.5$, and we subsequently take these as our fiducial values. They are lower than those inferred from the bootstrapping above for the two brighter bins, but consistent within the 1$\sigma$ uncertainties. This meets our expectation that the bootstrapped values are biased high in a regime of strong clustering and low number statistics. While this effect may also have an impact on the values inferred through MCMC-fitting, it is to some degree mitigated by our prior and likelihood choice, which forces the variance to be small unless the data strongly prefers it (see Section \ref{sec:mcmc_fitting}). The raw posterior samples of the mean, standard deviation, and inferred $\sigma_\mathrm{CV}$ are shown in Figure \ref{fig:cv_posteriors} in Appendix \ref{sec:cv_posteriors}.

\begin{figure}
     \centering
     \includegraphics[width=0.47\textwidth]{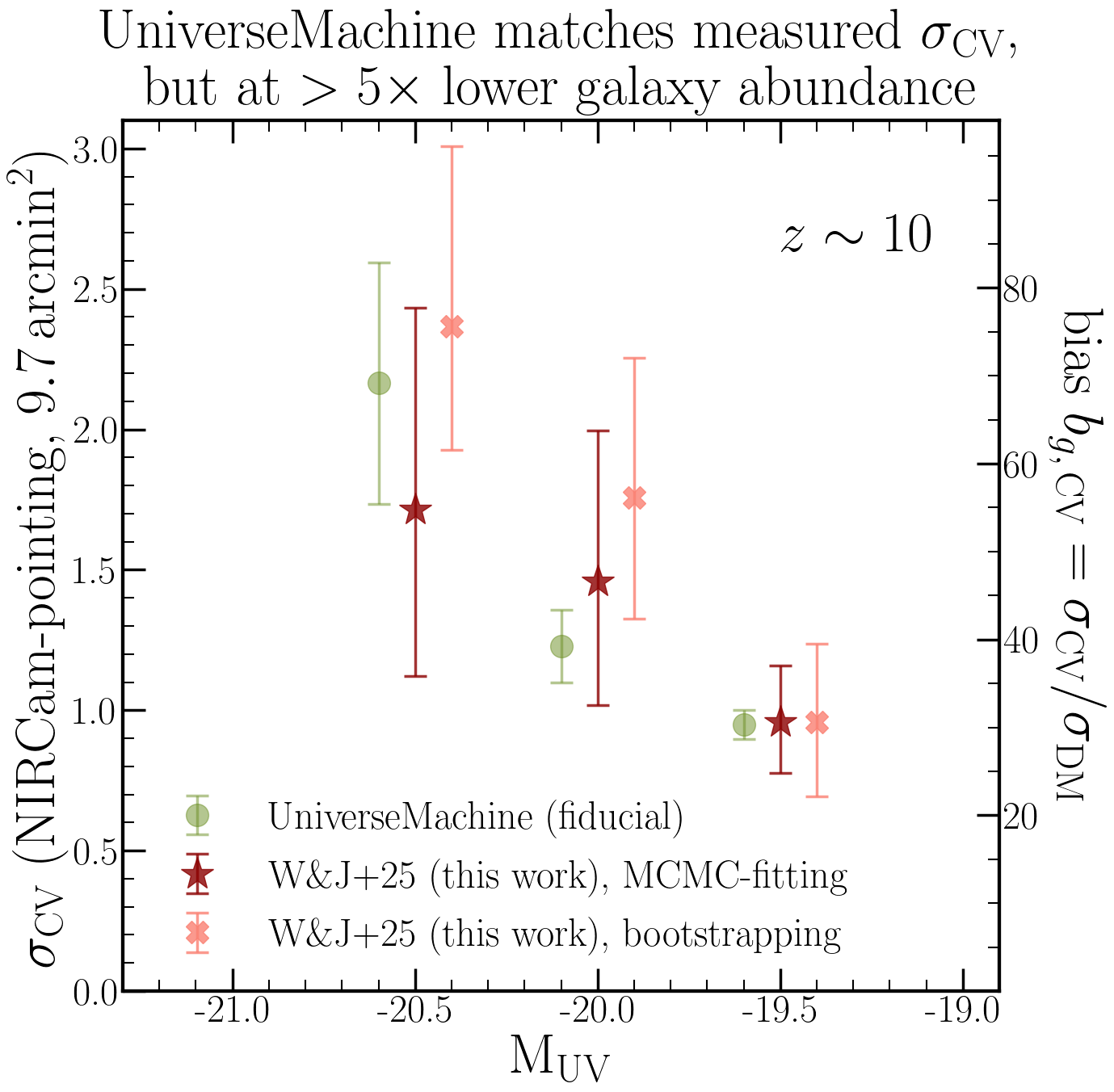}
     \caption{Cosmic variance $\sigma_{\rm CV}$ in the galaxy number count at $z\sim10$ for a NIRCam pointing sized survey ($9.7\,{\rm arcmin}^2$). We plot values inferred through two different methods: bootstrapping Equation \ref{eq:cv_bootstrap} over 34 independent fields (crosses), and MCMC-fitting to the distribution of number counts per field (stars). The secondary y-axis shows the galaxy bias $b_{g,\,\rm CV}$, inferred from $\sigma_{\rm CV}$ using Equation \ref{eq:bias_cv}. Measurements are shown for three M$_{\rm UV}$ limits, M$_{\rm UV}<-19.5$, $-20$, and $-20.5$ with different markers being slightly displaced on the x-axis for better visual separation. For comparison, we show values extracted using the same MCMC-fitting method from the \texttt{UniverseMachine} (green dots). While they match our measurements in all three bins of M$_{\rm UV}$, the \texttt{UniverseMachine} under-predicts the abundance of UV-bright galaxies at $z\sim10$ by a factor $\gtrsim5$ (see Figures \ref{fig:toy_models1} and \ref{fig:toy_models2}).}
     \label{fig:cosmic_variance}
\end{figure}

In Figure \ref{fig:cosmic_variance}, we show the cosmic variance values inferred with the two different methods and for the three M$_{\rm UV}$ limits, and compare them to values from the \texttt{UniverseMachine}, measured using the MCMC-fitting approach. The secondary y-axis represents the cosmic variance inferred galaxy bias as defined in Equation \ref{eq:bias_cv}. The values from the \texttt{UniverseMachine} agree well with our measurements. Given that they are measured in a consistent way, this indicates that the clustering of UV-bright galaxies at $z\sim10$ in the \texttt{UniverseMachine} is realistic, and that it was reasonable to calibrate $\sigma_{\rm CV}$ to it in \citetalias{Weibel2026}. We note that at the same time, the \texttt{UniverseMachine} under-predicts the abundance of UV-bright galaxies at $z\sim10$ by a factor $\gtrsim5$ and further discuss the implications of this in Section \ref{sec:toy_model_comparison}. 

Our measured values of $\sigma_{\rm CV}$ imply that cosmic variance has a large impact on the uncertainty in the number counts of UV-bright galaxies at $z\sim10$. For example, at M$_{\rm UV}<-20.5$, we measure $\sigma_{\rm CV}=1.71^{+0.72}_{-0.59}$, suggesting that the cosmic variance driven fractional uncertainty in the galaxy number count inferred from a single NIRCam pointing is as high as $\sim110-240\,$\%. Given that the Poisson contribution to the uncertainty is $\sim\sqrt{N}$, this means that cosmic variance dominates the uncertainty in the number count by far in this case. The impact of cosmic variance scales with the number of independent fields as $\sigma_{\rm CV}\propto1/\sqrt{N_{\rm fields}}$, and it also decreases for larger field sizes as we will quantify in Section \ref{sec:non_linear_scales}.

\subsection{Model Comparison}
\label{sec:toy_model_comparison}

Our measurements of cosmic variance and thus the clustering strength of LBGs at $z\sim10$ put complementary constraints on galaxy physics at cosmic dawn. The \texttt{UniverseMachine} reproduces the cosmic variance we measure (see Figure \ref{fig:cosmic_variance}), but it under-predicts the number density of UV-bright galaxies (i.e. the mean number of galaxies per NIRCam pointing) at $z\sim10$ by a factor of $4.9\pm0.8$, $5.3\pm1.2$ and $8.7\pm1.9$ for M$_{\rm UV}<-19.5$, $-20$, and $-20.5$, corresponding to 4.8, 3.6, and $4.2\sigma$ tensions between the simulation and the data respectively. Various models have been proposed in the literature to account for the observed abundance of galaxies at $z\sim10$. While they were all designed to reproduce the observed galaxy number density (see e.g., \citetalias{Weibel2026}), they affect the clustering in different ways, meaning that an accurate measurement of the clustering strength can be used to distinguish between different models. As discussed and quantified in e.g. \citet{Mirocha2020}, \citet{Munoz2023}, \citet{Shen2024}, and \citet{Gelli2024}, this is particularly true for models invoking stochasticity in the SFH to boost galaxy number counts. Such models increase the scatter in the M$_{\rm UV}$-$M_{\rm halo}$ relation, so that observed UV-bright galaxies tend to reside in lower mass halos that have up-scattered in UV-luminosity due to an ongoing starburst. Since lower mass halos are less clustered, this decreases the clustering signal. In addition, the halos that have up-scattered in M$_{\rm UV}$ at a given point in time are a random sub-set of all the halos of that mass, further diluting their clustering strength.

Here, we implement different models in the \texttt{UniverseMachine} and quantify their effect on both the number density, and the cosmic variance. Our simple models represent classes of models that have been proposed in the literature as an explanation for the high abundance of UV-bright galaxies at $z\sim10$. To facilitate comparing models throughout the paper, we assign a label consisting of a letter and a number to each of them. We provide a brief overview of the models here, and discuss them in more detail in the following sections.

\begin{figure*}
    \centering
        \includegraphics[width=\textwidth]{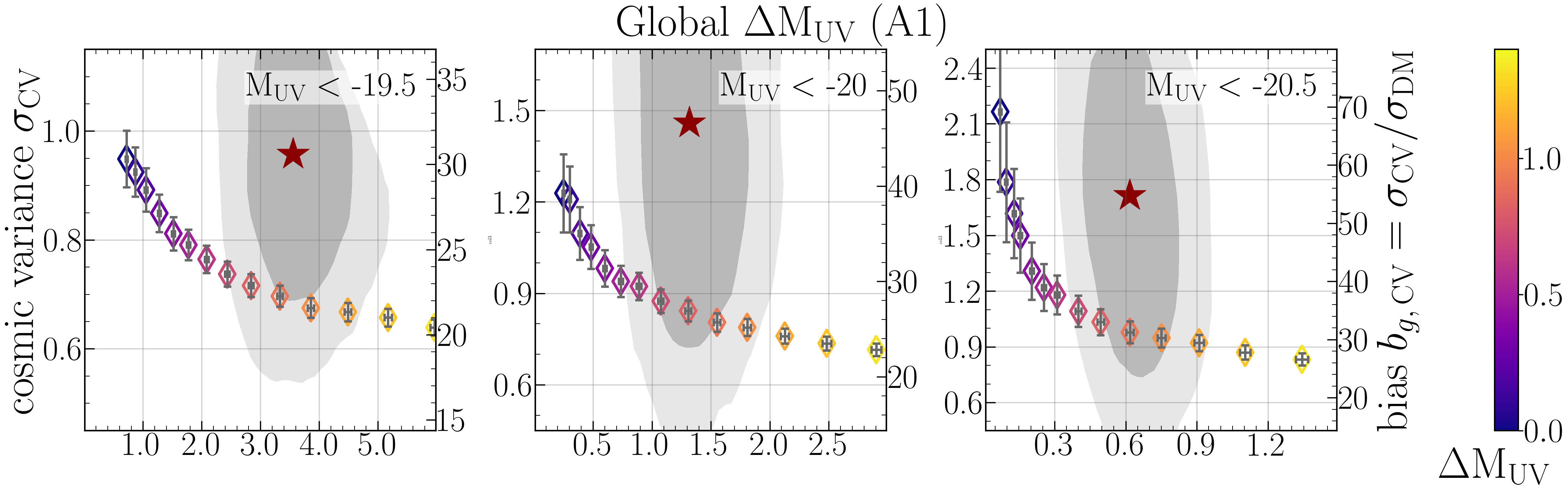}
        \includegraphics[width=\textwidth]{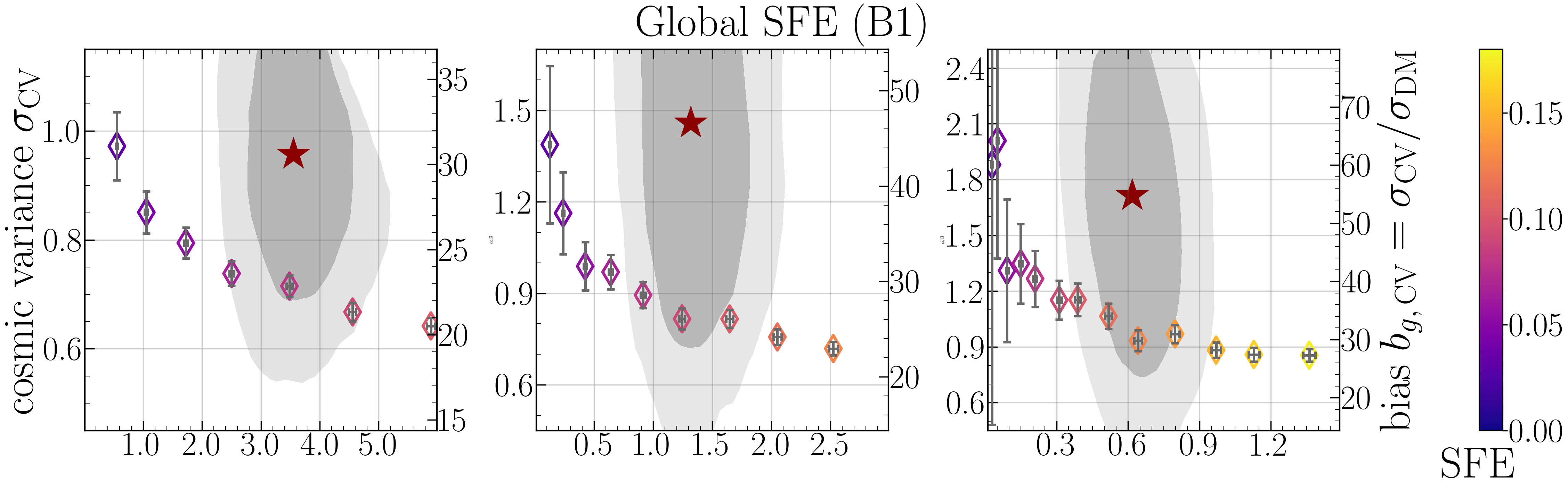}
        \includegraphics[width=\textwidth]{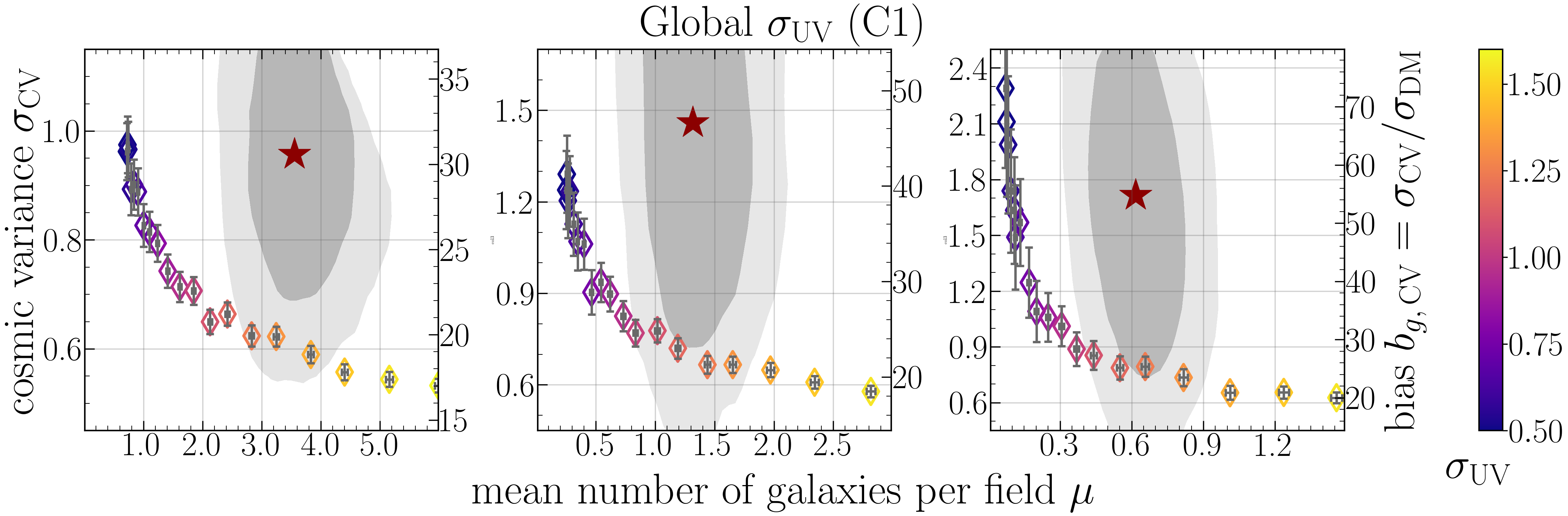}
    \caption{Combined constraints on the abundance of galaxies at $z\sim10$, quantified as the mean number of galaxies per field (i.e. per NIRCam pointing), $\mu$,  and their clustering, quantified by the cosmic variance $\sigma_{\rm CV}$. Each row of panels corresponds to a simple model implemented in the \texttt{UniverseMachine} to represent a class of models invoked to explain the abundance of UV-bright galaxies at $z\sim10$. From left to right, panels correspond to different M$_{\rm UV}$ limits as indicated in the top row of panels. The red stars and gray contours show our measurements, and the 1, and 2$\sigma$ confidence regions obtained through MCMC-fitting. We plot different realizations of each model by gradually increasing the quantity of interest: $\Delta {\rm M_{UV}}$, the boost to the UV-luminosity of all galaxies (in mag); the SFE (defined as $M_*/(f_b\,M_{\rm halo})$); and $\sigma_{\rm UV}$, the scatter in the M$_{\rm UV}$-M$_{\rm halo}$ relation (in mag). This illustrates that different models follow similar tracks on the $\sigma_{\rm CV}$-$\mu$-plane, showing an inverse relationship between $\mu$ and $\sigma_{\rm CV}$. Models can however be distinguished by considering different M$_{\rm UV}$ limits simultaneously (see Section \ref{sec:model_tension}).}
    \label{fig:toy_models1}
\end{figure*}

\begin{description}
    \item[Global $\mathbf{\Delta}$M$_\mathbf{UV}$ (A1)] We boost the UV-luminosity of all galaxies in the \texttt{UniverseMachine} by $\Delta {\rm M_{UV}}$ in magnitudes, without changing any other properties. This simulates a uniform decrease in the mass-to-light ratio across all galaxies, e.g. through a global change towards a more top-heavy IMF, or a lack of dust attenuation. 
    \item[$\mathbf{\Delta}$M$_\mathbf{UV}$ for $\mathbf{M_{halo}>X}$ (A2)] Instead of boosting the UV-luminosity of all galaxies, we only boost it for galaxies residing in halos with log($M_{\rm halo}/{\rm M_\odot})>10$, and $>10.5$. This simulates the effect of e.g., AGN boosting the UV-luminosity of galaxies that reside in massive halos.
    \item[Global SFE (B1)] We model a constant SFE across all galaxies in the \texttt{UniverseMachine}, where ${\rm SFE}=M_*/(f_b\,M_{\rm halo})$ and $f_b=0.16$ is the baryon fraction, and then continuously increase SFE to boost the number of UV-bright galaxies. In contrast to the boost in M$_{\rm UV}$, this model starts from $M_{\rm halo}$ in the \texttt{UniverseMachine} which we convert to $M_* = {\rm SFE}\,f_b\,M_{\rm halo}$. $M_*$ is then converted to M$_{\rm UV}$, assuming the typical $M_*$-${\rm M_{UV}}$-relation in the \texttt{UniverseMachine}, including scatter. 
    \item[DMSFE: $\mathbf{SFE\propto M_{halo}^q}$ (B2)] Building on the previous model, we implement a power-law scaling of the SFE with $M_{\rm halo}$ with a non-zero slope, ${\rm SFE}\propto M_{\rm halo}^{0.5}$, similar to the density modulated SFE (DMSFE) scenario proposed by \citet{Somerville2025}. To explore the impact of the slope of this scaling law, we also test a model with a sharper scaling, ${\rm SFE}\propto M_{\rm halo}^{0.6}$. In both cases, we continuously increase the normalization of the relation defined as ${\rm SFE_{peak}}={\rm SFE}(M_{\rm halo}=10^{11.6}\,{\rm M_\odot})$, the SFE at the maximum halo mass in the \texttt{UniverseMachine}. 
    \item[Global $\mathbf{\sigma_{UV}}$ (C1)] To mimic the effect of enhanced stochasticity in the star formation, we increase the global scatter in the M$_{\rm UV}$-$M_{\rm halo}$ relation, denoted as $\sigma_{\rm UV}$. It is important to note that $\sigma_{\rm UV}$ cannot be uniquely related to stochasticity in the SFHs of galaxies. Other processes that contribute to $\sigma_{\rm UV}$ are variations in the metallicity, dust attenuation, IMF, SFE, or the nebular contribution to the UV-continuum. 
    \item[$\mathbf{M_{halo}}$-dependent $\mathbf{\sigma_{UV}}$ (C2)] Finally, we implement a scaling of $\sigma_{\rm UV}$ with $M_{\rm halo}$ following \citet{Sun2023}, $\sigma_{\rm UV} = \sigma_{\rm UV,\,norm} - 1/3{\rm log(}M_{\rm halo}/10^{10}\,{\rm M_\odot})$, where $\sigma_{\rm UV,\,norm}=\sigma_{\rm UV}(M_{\rm halo}=10^{10}{\rm M_\odot})$. 
\end{description}

\begin{figure*}
    \centering
        \includegraphics[width=\textwidth]{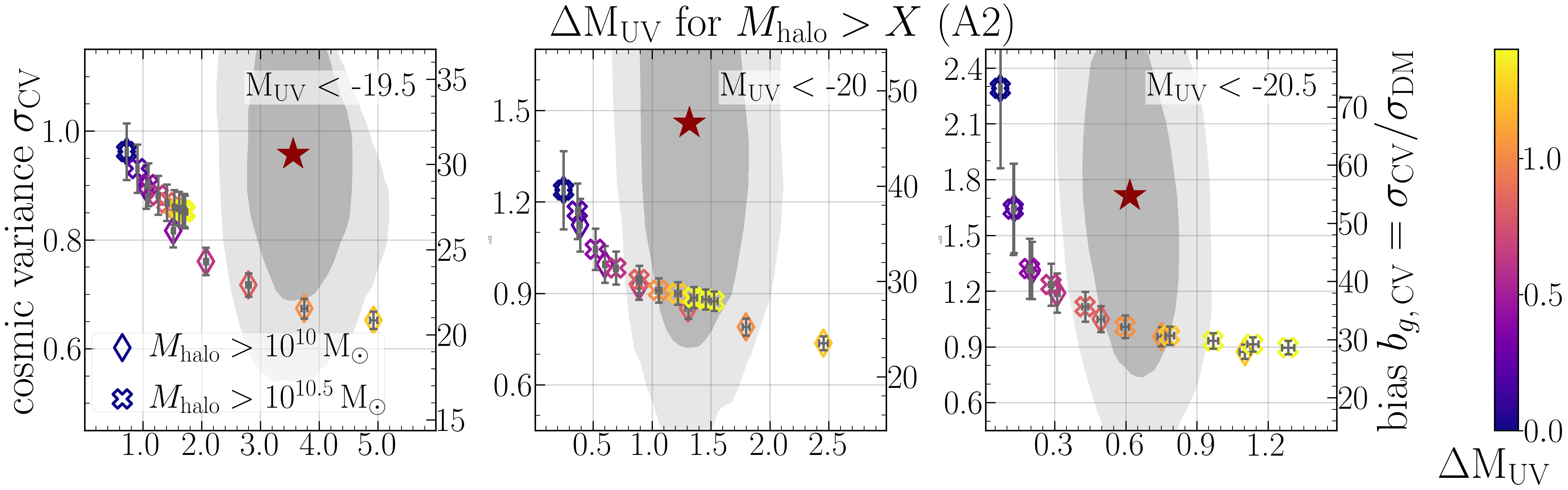}
        \includegraphics[width=\textwidth]{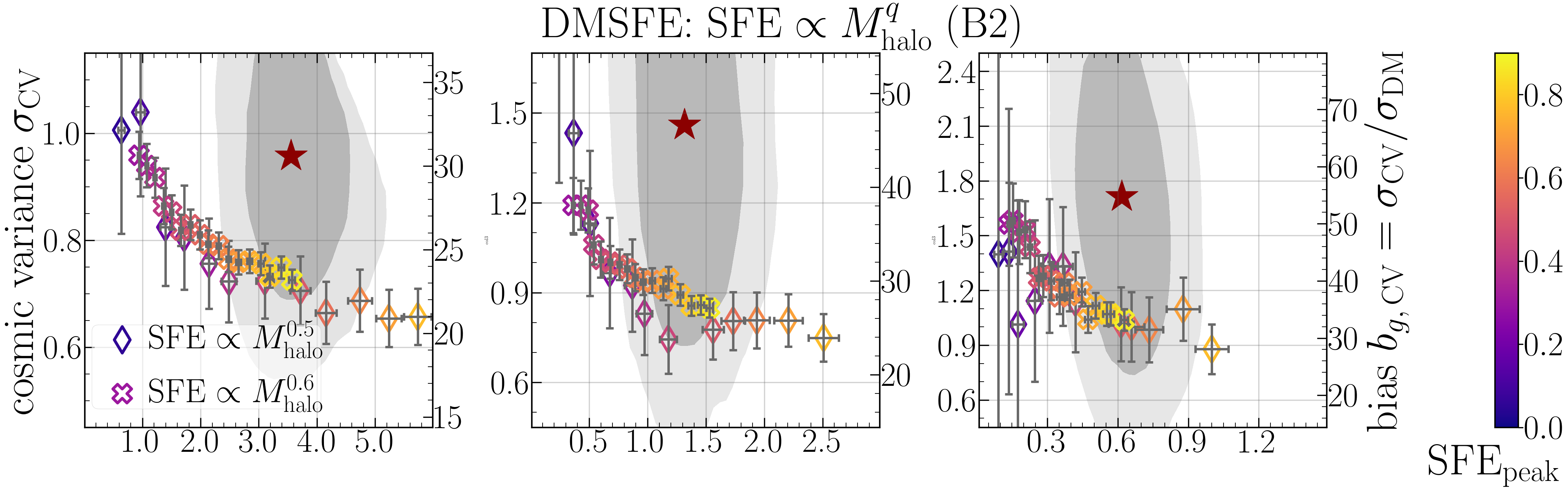}
        \includegraphics[width=\textwidth]{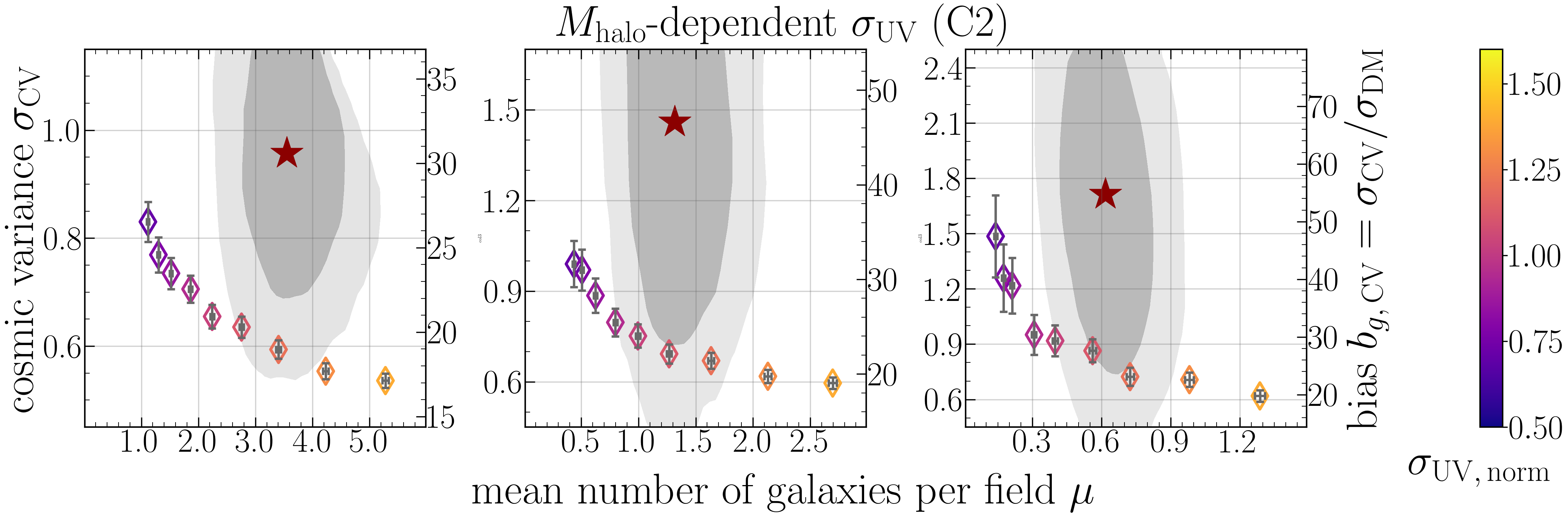}
    \caption{Same as Figure \ref{fig:toy_models1}, but for three models where the model parameters depend on halo mass. In the first one (top, A2), the UV-luminosity is boosted by $\Delta \rm M_{UV}$ in halos above a threshold mass. If the threshold mass is low enough ($10^{10}\,{\rm M_\odot}$ or lower), this produces results that are nearly identical to a global boost (Figure \ref{fig:toy_models1}, A1). If the threshold mass is higher $(>10^{10.5}\rm M_\odot)$, there are not enough halos of that mass to reproduce the measured number density of M$_{\rm UV}<-19.5$ galaxies. The second model (middle, B2) implements two power-law scalings of the SFE with halo mass, assuming two different slopes (0.5 and 0.6), akin to the density modulated SFE scenario from \citet{Somerville2025}. Both produce models consistent with the data within 1$\sigma$, with the steeper slope implying a higher SFE$_{\rm peak}$. The third model (bottom, C2) shows a scaling of $\sigma_{\rm UV}$ with log($M_{\rm halo}$), following \citet{Sun2023}. Similar to the global increase in $\sigma_{\rm UV}$, this is inconsistent with the data at $>1\sigma$ for each M$_{\rm UV}$ limit individually.}
    \label{fig:toy_models2}
\end{figure*}

For each model, we create multiple realizations, gradually increasing the quantity of interest ($\Delta M_{\rm UV}$, SFE, SFE$_{\rm peak}$, $\sigma_{\rm UV}$, and $\sigma_{\rm UV,\,norm}$ respectively), and apply the MCMC-fitting explained in Section \ref{sec:mcmc_fitting} to infer the mean number of galaxies per field $\mu$, and $\sigma_{\rm CV}$. We present the resulting combined constraints on the abundance and the cosmic variance in Figures \ref{fig:toy_models1} and \ref{fig:toy_models2}. Each set of three panels represents our three M$_{\rm UV}$ limits. The contours show the 1$\sigma$ and 2$\sigma$ confidence regions of the MCMC-fitting to the data from Section \ref{sec:mcmc_results} (identical to the bottom left panels in Figure \ref{fig:cv_posteriors}). 

\subsubsection{General Observations}
\label{sec:toy_models_general}

The tracks that the different models describe on the $\sigma_{\rm CV}-\mu$ plane look overall similar as they show the same qualitative trend of decreasing $\sigma_{\rm CV}$ for higher $\mu$. This is because all models increase the normalization of the M$_{\rm UV}$-$M_{\rm halo}$ relation such that at fixed halo mass, galaxies have a higher UV-luminosity. Conversely, this means that galaxies at fixed M$_{\rm UV}$ reside in lower mass halos which are less clustered. This is true for any model that increases the number density of UV-bright galaxies without changing the properties of the underlying dark matter halos. In order to increase the number of bright galaxies, they have to reside in halos that are more abundant, i.e. in halos of lower mass. As a consequence, none of the explored models is capable of exactly reproducing our combined measurements of the abundance and the clustering strength shown by a red star in Figures \ref{fig:toy_models1} and \ref{fig:toy_models2}. We quantify the statistical significance of the resulting tension in Section \ref{sec:model_tension}, but note here that it may also indicate that our measurements are still biased high due to the reasons outlined in Section \ref{sec:bootstrapping_results} (see also \citealt{Jespersen2025}).

While the different models show a qualitatively similar behavior, they differ in their detailed quantitative predictions. For example, models that increase $\sigma_{\rm UV}$ decrease $\sigma_{\rm CV}$ most significantly for all M$_{\rm UV}$ limits. It is further important to note that in order for a model to successfully reproduce our measurements, it has to match the constraints for all three M$_{\rm UV}$ limits simultaneously (see Section \ref{sec:model_tension}). Subsequently, we discuss the different models shown in Figures \ref{fig:toy_models1} and \ref{fig:toy_models2}.

\subsubsection{Models Changing Global Properties}
\label{sec:toy_models_global}

We start with the models in Figure \ref{fig:toy_models1}: a global boost in M$_{\rm UV}$ (model A1), a global increase in the SFE (B1), and enhanced scatter in the M$_{\rm UV}$-$M_{\rm halo}$ relation, independent of halo mass (C1). Boosting the UV-luminosity of all galaxies by $\Delta {\rm M_{UV}}\sim1\,$mag matches the number density constraints in all three panels, while the implied $\sigma_{\rm CV}$ is marginally consistent with our measurements within $1\sigma$ respectively. The same is true for a model with $\sigma_{\rm UV}\sim1.2$, although slightly more scatter is required to match the higher number density at M$_{\rm UV}<-19.5$, and the respective models are all inconsistent with our clustering measurements at $\gtrsim1\sigma$ significance across the three magnitude limits. The constant SFE model in turn struggles to reproduce the number densities for all three M$_{\rm UV}$ limits simultaneously. At M$_{\rm UV}<-19.5$, an SFE of $\sim0.1$ is favored by the data, and a higher SFE quickly over-produces the number density of galaxies. However, an SFE of $\gtrsim0.15$ is needed to match the measured number density of M$_{\rm UV}<-20.5$ galaxies. In other words, a global increase in the SFE steepens the faint end slope of the UVLF. While we only probe the UVLF down to ${\rm M_{UV}\sim-19.5}$ here, the faint end slope has been measured down to ${\rm M_{UV}\sim-12.5}$ in e.g., \citet{Chemerynska26}, showing no steepening with respect to the measurements presented in \citetalias{Weibel2026}.

\subsubsection{Mass-Dependent Models}
\label{sec:toy_models_mass_dependent}

Recently proposed models to explain the observed galaxy abundance at $z\sim10$ often go beyond changing galaxy properties globally. They instead suggest a dependence of the SFE, $\sigma_{\rm UV}$, or the change in mass-to-light ratio on other physical properties. In Figure \ref{fig:toy_models2}, we first show the model where we only boost M$_{\rm UV}$ for galaxies residing in halos above two different threshold halo masses (model A2). For a mass threshold of log($M_{\rm halo}/{\rm M_\odot})>10$, this only marginally increases the cosmic variance at fixed mean number count compared to the global boost in M$_{\rm UV}$, and the same is true for any lower mass threshold. If we instead boost the UV-luminosity in halos with log($M_{\rm halo}/{\rm M_\odot})>10.5$, it is impossible to reproduce the number counts of galaxies with $M_{\rm UV}<-19.5$ because the number density of halos above the mass threshold is too low. Further, different boosts would be required to match the constraints for $M_{\rm UV}<-20$ and $M_{\rm UV}<-20.5$ respectively, making this model unfeasible. 

\begin{figure*}
    \centering
        \includegraphics[width=\textwidth]{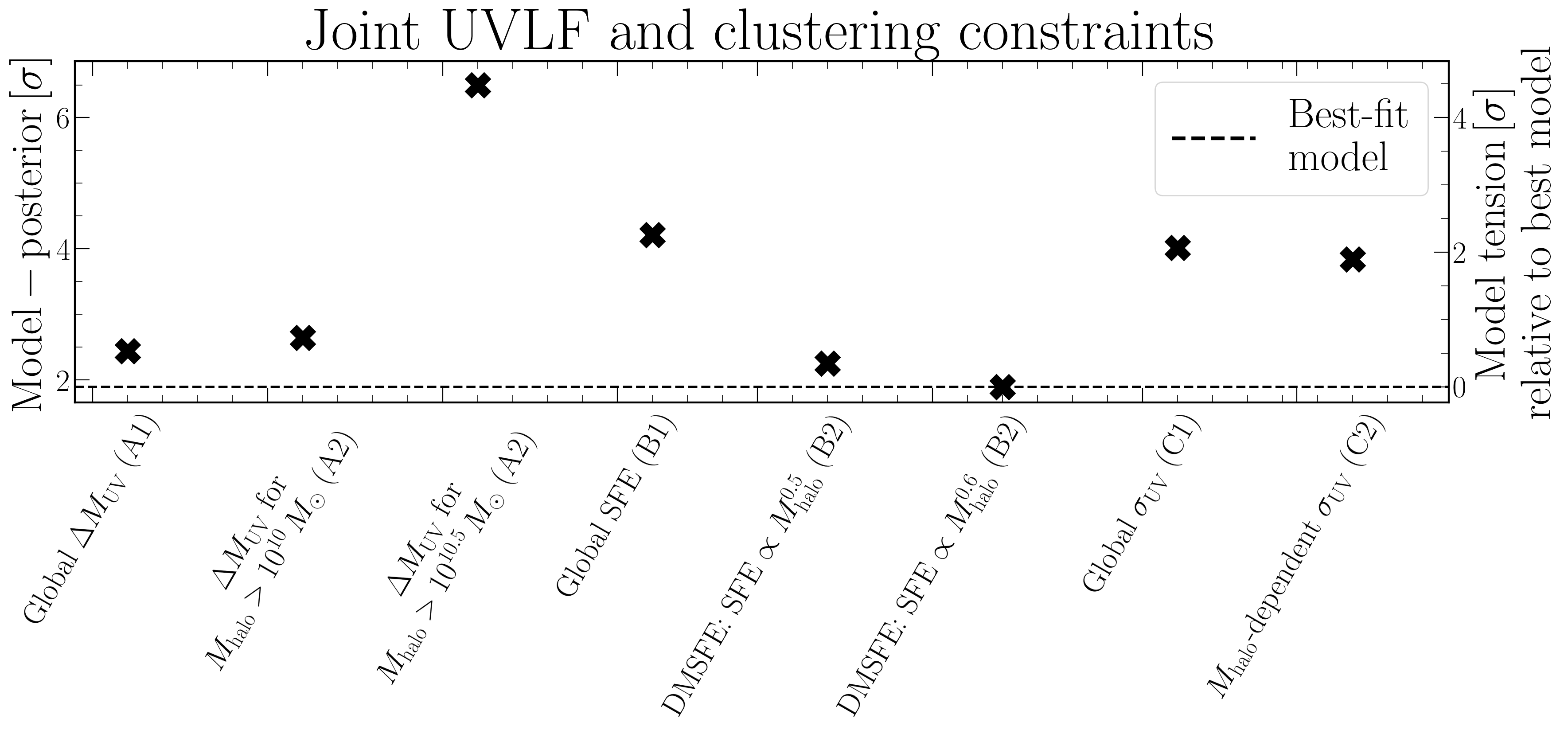}
    \caption{Combined model tension, calculated as the equivalent posterior distance in Gaussian units, across three distinct M$_{\rm UV}$ bins ($-20<{\rm M_{UV}}<-19.5$, $-20.5<{\rm M_{UV}}<-20$, and ${\rm M_{UV}}<-20.5$), for the different models presented in Section \ref{sec:toy_model_comparison}. The secondary y-axis on the right shows the combined model tension relative to the best-fitting model (sharp DMSFE), indicated as the horizontal dashed line. This shows that models that enhance the UV-scatter $\sigma_{\rm UV}$ (bursty SF) are disfavored at $\sim2\sigma$. The global increase in SFE is disfavored at $>2\sigma$, consistent with our qualitative assessment in Section \ref{sec:toy_models_global}. As can be clearly seen from Figure \ref{fig:toy_models2}, only boosting M$_{\rm UV}$ in halos with $M_{\rm halo}>10^{10.5}\,{\rm M_\odot}$ cannot reproduce the measured galaxy abundance, resulting in a highly disfavored model ($\Delta\sigma>4$).}
    \label{fig:model_tension}
\end{figure*}

Models where the SFE shows a power-law dependence on halo mass ${\rm SFE}\propto M_{\rm halo}^q$ (B2) only marginally enhance $\sigma_{\rm CV}$ at fixed $\mu$ compared to the increase in SFE at all halo masses (B1). However, as opposed to the global increase in SFE, they lie within the $1\sigma$ contours for all three M$_{\rm UV}$ limits simultaneously. The sharper scaling (${\rm SFE}\propto M_{\rm halo}^{0.6}$) provides a slightly better fit (see Section \ref{sec:model_tension}), but implies a higher peak efficiency, ${\rm SFE_{peak}}\sim0.94$. Even sharper slopes imply ${\rm SFE_{peak}}>1$ and are therefore un-physical. For reference, in the fiducial \texttt{UniverseMachine}, SFE$_{\rm peak}$ does not exceed 0.06.

Finally, analyzing the FIRE-2 simulation, \citet{Sun2023} found that the UV-scatter scales with halo mass because galaxies residing in lower mass halos exhibit more burstiness in their star formation. We show the model that assumes their proposed scaling in the bottom panels in Figure \ref{fig:toy_models2} (C2). Similar to the global increase in $\sigma_{\rm UV}$, these models are inconsistent with our measurements at $>1\sigma$ for all three M$_{\rm UV}$ limits individually. A normalization of the scaling law at $M_{\rm halo}=10^{10}\,{\rm M_\odot}$ of $\sigma_{\rm UV,\,norm}\sim1.1$ is required to match our number density constraints. We have also tested shallower and sharper scalings of $\sigma_{\rm UV}$ with log($M_{\rm halo}$), adopting proportionality constants of 1/4 and 1/2, and found no significant differences in terms of the resulting $\sigma_{\rm CV}$ and $\mu$. We specify the best-fitting model parameters for each model in Table \ref{tab:model_params} in Appendix \ref{sec:appendix_best_fit_parameters}.

\subsection{Combined Model Tension}
\label{sec:model_tension}

As mentioned above, a successful model has to match the observational constraints across the different M$_{\rm UV}$ limits. To quantify the combined tension between each model and the data, we repeat the measurements presented in Figures \ref{fig:toy_models1} and \ref{fig:toy_models2} in distinct M$_{\rm UV}$ bins ($-20<{\rm M_{UV}}<-19.5$, $-20.5<{\rm M_{UV}}<-20$, and ${\rm M_{UV}}<-20.5$). This is to ensure that each galaxy contributes only to one bin. In the case of M$_{\rm UV}$ limits, the brighter galaxies are present in multiple bins, introducing covariance between them. We show our measurements and model trajectories in analogy to Figures \ref{fig:toy_models1} and \ref{fig:toy_models2}, but in M$_{\rm UV}$ bins in Appendix \ref{sec:appendix_model_comp_bins}. The distinct bins allow us to treat bins independently and compute a combined model tension $m$ as the equivalent posterior distance in Gaussian units as $(\mu_{\rm model} - \mu_{\rm data})(\Sigma_{\rm model}+\Sigma_{\rm data})^{-1}(\mu_{\rm model} - \mu_{\rm data})^T$, where $\Sigma_{\rm model}$ and $\Sigma_{\rm data}$ are the covariance matrices for the data and the model constraints respectively. We show the combined model tension for all the models in Figure \ref{fig:model_tension}. At face value, all displayed models are at $\gtrsim2\sigma$ tension with the data. However, as discussed in Sections \ref{sec:mcmc_fitting} and \ref{sec:bootstrapping_results}, our measured $\sigma_{\rm CV}$ may be biased high and more data is needed to corroborate our measurements. We therefore subsequently focus on \textit{distinguishing} between different models based on their relative tension with the data.

Specifically, we compare the tension of all models to the model that shows the best fit, which is the sharp DMSFE model (B2, ${\rm SFE}\propto M_{\rm halo}^{0.6}$). Consistent with our qualitative assessment above, the only model that is disfavored at a high ($>4\sigma$) confidence is the one that boosts M$_{\rm UV}$ for log($M_{\rm halo}/{\rm M_\odot})>10.5$ (A2). Further, models that enhance the UV-scatter $\sigma_{\rm UV}$ (C1, C2) are tentatively disfavored at $\sim2\sigma$ significance relative to the sharp DMSFE model, and a global increase in the SFE (B1) is disfavored at $>2\sigma$. While these differences are not significant enough to draw any robust conclusions, we show in Section \ref{sec:future_prospects} that with sufficient additional imaging along independent lines of sight, this method allows for a robust distinction between different models.

\section{Discussion}
\label{sec:discussion}

Using a sample of LBGs at $z\sim10$, compiled from imaging along 34 independent lines of sight in \citetalias{Weibel2026}, we performed the first direct measurement of $\sigma_{\rm CV}$ at such high redshifts and found that the cosmic variance per NIRCam pointing is $\gtrsim100$\%. Relating $\sigma_{\rm CV}$ to galaxy clustering, we illustrated that different models proposed to explain the abundance of UV-bright galaxies at $z\sim10$ can, in principle, be distinguished based on combined constraints on $\sigma_{\rm CV}$ and the UVLF. Subsequently, we quantify the impact of clustering on non-linear scales on our measured $\sigma_{\rm CV}$, and its implications for comparisons to measurements of the linear galaxy bias. We further discuss remaining caveats and future prospects for clustering measurements based on pure parallel imaging data. 

\subsection{Effect of Non-Linear Scales}
\label{sec:non_linear_scales}

By definition, the field-to-field variance is affected by clustering on all scales smaller than the survey volume. On small scales, the clustering of galaxies is enhanced by non-linear effects. The situation is complicated by the non-trivial survey geometry considered here. A NIRCam pointing consists of two quadratic modules at a small separation of 44\arcsec. It can therefore be approximated by a rectangle of dimensions $4.4\,$\arcmin$\times\,2.2\,$\arcmin. By testing in the \texttt{UniverseMachine}, we have verified that the effect of considering the gap between modules is of the order of a few percent, many times smaller than the uncertainty on even the cosmic variance as measured in the simulation. In practice, the third dimension of the survey volume is determined by the selection function in redshift space which we approximate by a top-hat function in the range $9.2<z<10.9$. The resulting survey volume is a cone whose physical size is much larger in the dimension pointing away from the observer than on the sky. An angular separation of $4.4\,$\arcmin\ on the sky corresponds to $1.1\,$pMpc at $z=10$ while the physical distance between a galaxy at $z=9.2$ and a galaxy at $z=10.9$ (assuming no angular separation) is $\sim34\,$pMpc. With this survey geometry, larger scales are therefore only sampled along a narrow cone while smaller scales are sampled in all directions. 

To quantify the contribution of clustering on small, non-linear scales to the cosmic variance, (and therefore to the galaxy bias $b_{g,\,\rm CV}$, measured from $\sigma_{\rm CV}$ using Equation \ref{eq:bias_cv}), we compute $b_{g,\,\rm CV}$ from the \texttt{UniverseMachine} in the same way as explained above, but for different field sizes. For comparison, we also compute the linear bias. Knowing both the halo mass, and the number of galaxies at a given M$_{\rm UV}$ that reside in each halo in the \texttt{UniverseMachine}, we can apply Equation (15) in \citet{Paquereau2025} to infer the linear galaxy bias as a number-weighted average of the linear halo bias from \citet{Tinker2010}.

\begin{figure}
     \centering
     \includegraphics[width=0.47\textwidth]{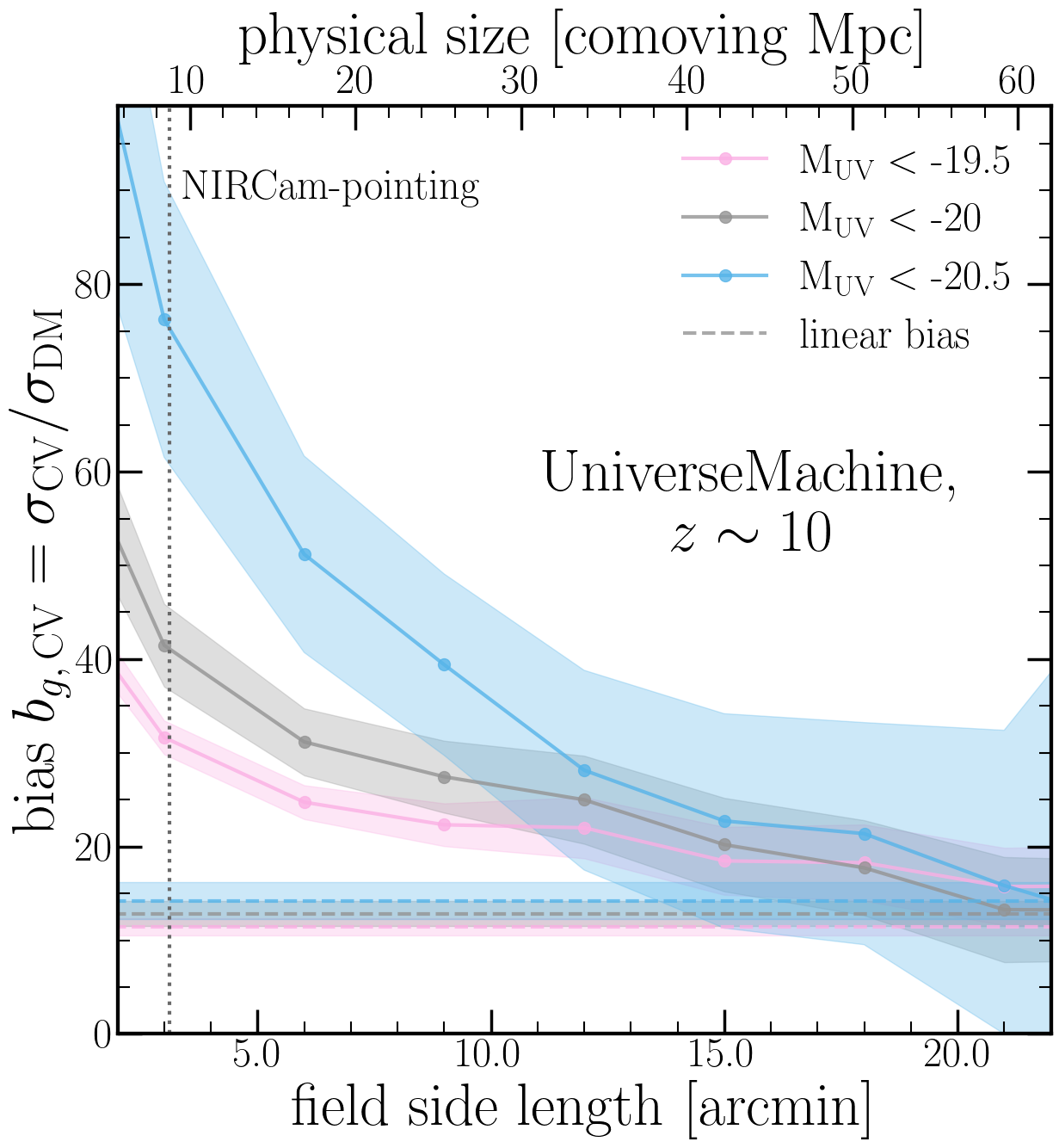}
     \caption{Galaxy bias $b_{g,\,\rm CV}$, as defined in Equation \ref{eq:bias_cv}, as a function of the field side length in arcmin for M$_{\rm UV}<-19.5$, -20, and -20.5, measured from the \texttt{UniverseMachine} at $z\sim10$. The secondary x-axis corresponds to the physical scale for the respective field side length in co-moving Mpc at $z=10$. Linear bias values are shown as horizontal dashed lines, and the shaded regions are 1$\sigma$ uncertainties respectively. This illustrates how the measured bias increases for small field sizes and exceeds the linear bias, driven by an increasing contribution from clustering on small, non-linear scales. If approximated by a square, a NIRCam pointing corresponds to a field side length of $\sim3.1\,$arcmin which is shown as a vertical dotted line where the bias deviates strongly from the linear value and shows a clear dependence on M$_{\rm UV}$.}
     \label{fig:field_sizes}
 \end{figure}

 We show the results of this exercise in Figure \ref{fig:field_sizes} where we plot $b_{g,\,\rm CV}$ as a function of the field side length for our three different M$_{\rm UV}$ limits, along with the scale-independent linear bias respectively. The bias inferred from $\sigma_{\rm CV}$ strongly increases for small field sizes, and exceeds the linear bias, driven by an increasing contribution of clustering on small, non-linear scales with the effect being stronger for brighter galaxies. At the size of a NIRCam pointing (field side length of $\sim3.1\,$arcmin), $b_{g,\,\rm CV}$ is a factor of $3-5$ higher than the linear bias, with a strong dependence on the M$_{\rm UV}$ limit (see Section \ref{sec:comparison}). For field side lengths $\gtrsim20\,$arcmin (corresponding to physical sizes $\gtrsim60\,$cMpc as shown on the secondary x-axis), $b_{g,\,\rm CV}$ approaches the linear bias value as the contribution of small-scale clustering becomes negligible. This is associated with larger uncertainties due to the limited size of the individual light cones in the \texttt{UniverseMachine}. Nevertheless, this illustrates the general trend of $b_{g,\,\rm CV}$ with field size, and its convergence to the linear bias provided sufficiently large fields. In principle, Figure \ref{fig:field_sizes} can be used to convert the measured $b_{g,\,\rm CV}$ to a linear bias, and compare it to inferred linear bias values in the literature, as we will do in the following.

\subsection{Comparing to Measurements and Models of the Linear Bias}
\label{sec:comparison}

We use the ratio between $b_{g,\,\rm CV}$ for a NIRCam pointing-sized field, and the linear bias in the \texttt{UniverseMachine} to scale our measured $b_{g,\,\rm CV}$ to the corresponding linear bias. For M$_{\rm UV}<-19.5$, $-20$, and $-20.5$, the conversion factors can be inferred along the vertical dotted line in Figure \ref{fig:field_sizes} to be $2.78^{+0.28}_{-0.26}$, $3.24^{+0.49}_{-0.47}$, and $5.37^{+1.29}_{-1.27}$. Dividing our fiducial measurements of $b_{g,\,\rm CV}$ by these factors, we get $b_{g,\,\rm CV,\,corr.}=11.0_{-2.3}^{+2.6}$, $14.4_{-4.8}^{+5.7}$, and $10.2_{-4.3}^{+4.9}$. These values are consistent with each other within 1$\sigma$, and show no obvious trend with M$_{\rm UV}$.

Measurements of the linear galaxy bias at $z\sim10$ have only become available recently. \citet{Dalmasso2024} used imaging from JADES to measure the 2PCF at $5\leq z\leq 11$, finding linear bias values of $b_{g,\,\rm lin.,\,D+24}=7.2\pm1.5$, and $9.6\pm1.7$ for $9<z<10$ and $10<z<11$. These are somewhat lower than our corrected bias values, which may be due to their selection of galaxies with M$_{\rm UV}<-17$, much fainter than the galaxies studied here. \citet{Paquereau2025} instead derived 2PCFs across $0.1< z <12$ from COSMOS-Web data \citep{Casey2023} above different stellar mass thresholds. Using the $M_*$-M$_{\rm UV}$ relation from \citet{Song2016}, our M$_{\rm UV}$ limits roughly correspond to mass thresholds of log$(M_*/{\rm M_\odot})>8.2$, 8.5 and 8.7. We therefore compare to their bias measurement for log$(M_*/{\rm M_\odot})>8.5$ which is available at $z\sim9.25$ as $b_{g,\,\rm lin.,\,P+25}=10.34_{-0.47}^{+0.69}$, consistent with our corrected bias values for all three M$_{\rm UV}$ limits. 
 
\citet{Gelli2024} used a simple analytical model to quantify the effect of stochastic star formation on the bias. They implemented mass dependent scatter in the M$_{\rm UV}$-M$_{\rm halo}$-relation following \citet{Sun2023}, and computed linear bias values for different M$_{\rm UV}$ limits, with and without UV scatter. We consider their model for M$_{\rm UV}<-19.5$, and the corresponding bias values at $z\sim9.8$, which are $b_{g,\,\rm lin.,\,G+24}=11.73$, and $b_{g,\,\rm lin.,\,G+24,\,bursty}=9.39$. Our corrected bias value for the same M$_{\rm UV}$ limit of $b_{g,\,\rm CV,\,corr.}({\rm M_{UV}}<-19.5)=11.0_{-2.3}^{+2.6}$ is close to their value without UV-scatter, but still consistent with the bursty value, emphasizing that more data is needed to distinguish between such models.

These comparisons illustrate that using the \texttt{UniverseMachine} to factor out the effect of clustering on non-linear scales yields linear bias values consistent with measurements in the literature. While the current uncertainty in $\sigma_{\rm CV}$ is too large to allow for a distinction between different models, it is important to note that the uncertainties that are specified for linear bias measurements from fitting an HOD to the 2PCF are artificially shrunk because they are calculated within the framework of a simplified model. On the other hand, $\sigma_{\rm CV}$ probes both the clustering of halos as well as the physical processes that determine the properties of the galaxies within them. This complicates its interpretation but it also means that $\sigma_{\rm CV}$ may have more constraining power in terms of discriminating between physical models, as we further explore below.

\begin{figure*}
    \centering
    \includegraphics[width=0.8\textwidth]{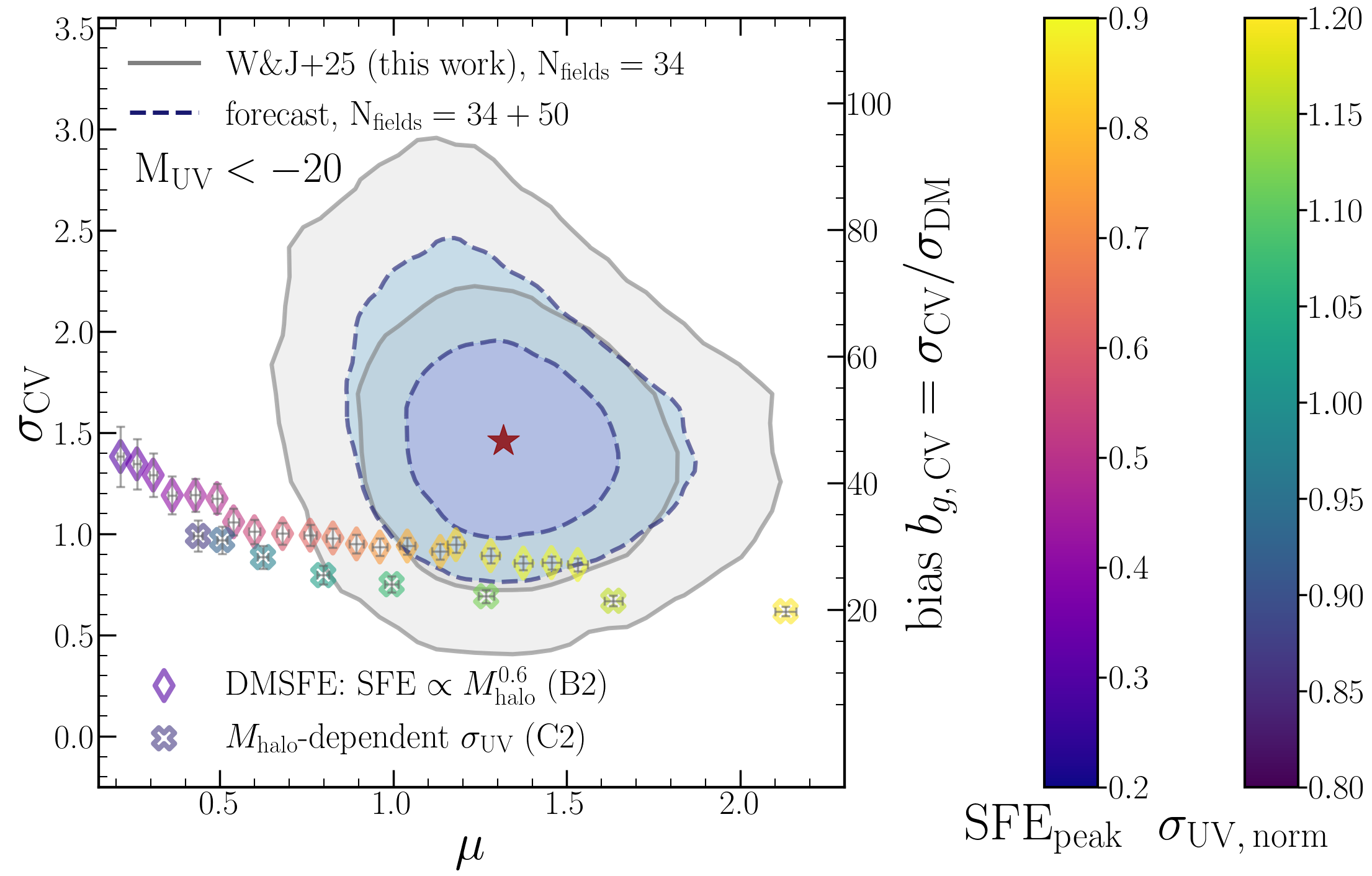}
    \caption{Combined constraints $\mu$ and $\sigma_{\rm CV}$ in analogy to Figures \ref{fig:toy_models1} and \ref{fig:toy_models2}. The gray contours and red star represent our measurement in the M$_{\rm UV}<-20$ bin. The blue contours are derived assuming that our measurements represent the ground truth and simulating the constraining power of 50 \textit{additional} lines of sight, as could be easily obtained in a dedicated pure parallel NIRCam imaging program. The colored markers represent our models with $M_{\rm halo}$-dependent UV-scatter (C2), as well as the DMSFE with ${\rm SFE}\propto M_{\rm halo}^{0.6}$ (B2). While our current constraints only disfavor such models at $\sim1\sigma$ confidence, they could be ruled out at $>2\sigma$ in a single M$_{\rm UV}$ bin with 50 additional independent NIRCam pointings.}
    \label{fig:future}
\end{figure*}

\subsection{Future Prospects for Measurements of Cosmic Variance}
\label{sec:future_prospects}

The uncertainty in the measured $\sigma_{\rm CV}$ roughly scales with the square root of the available number of independent lines of sight. A substantial improvement in the bias measurement is therefore in reach with \textit{JWST}/NIRCam pure parallel imaging, which is the perfect observing mode to acquire imaging data in random directions in the sky at very low overhead costs. 

In Figure \ref{fig:future}, we show the hypothetical 1, and 2$\sigma$ contours on the $\sigma_{\rm CV}$-$\mu$-plane at M$_{\rm UV}<-20$ which we could obtain with only 50 additional lines of sight, assuming the depth distribution of the 34 fields used in this work. We again plot the model with halo mass dependent UV-scatter \citep[][, C2]{Sun2023}, as well as the DMSFE model \citep[][, B2]{Somerville2025} already shown in Figure \ref{fig:toy_models2}. While models that match the measured $\mu$ are only disfavored at $\sim1\sigma$ confidence with the current data, they could already be disfavored at $>2\sigma$ confidence with 84 independent lines of sight ($34 + 50$). Such a data set could easily be obtained in a dedicated pure parallel imaging program in just one observational cycle \citep[e.g.,][]{jwpure}, and will likely become available through a combination of parallel as well as primary imaging programs across observational cycles in the future. 

\begin{figure*}
    \centering
        \includegraphics[width=\textwidth]{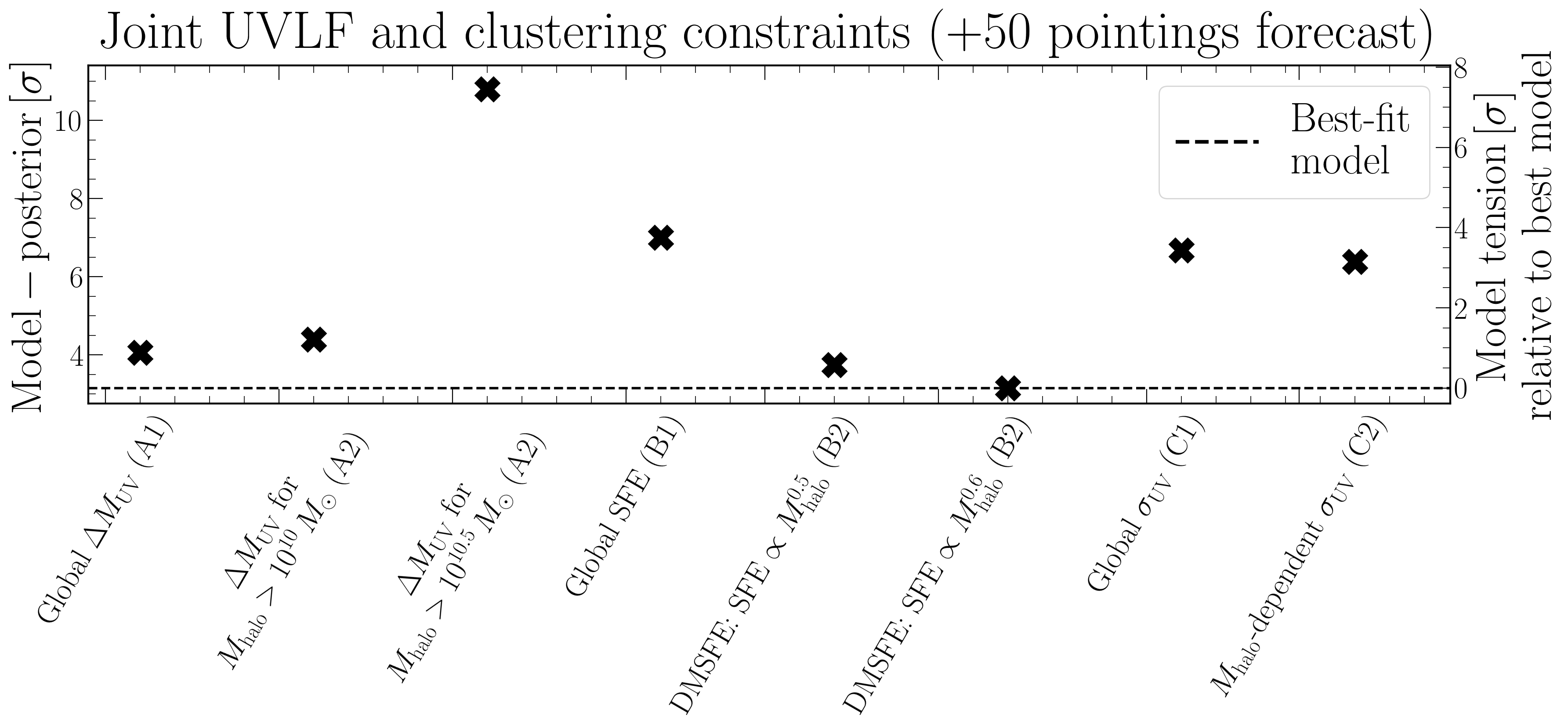}
    \caption{Same as Figure \ref{fig:model_tension}, but with 50 additional lines of sight following the depth distribution of the 34 fields used in this work. This shows that with only 50 additional independent NIRCam pointings, different models can be distinguished at statistically significant levels based on joint UVLF and clustering constraints. In particular, models that invoke enhanced UV-scatter (C1, C2), as well as a global increase in the SFE (B1) can be ruled out at $>3\sigma$, relative to DMSFE models (B2), and models that boost the UV-luminosity of galaxies (A1, A2).}
    \label{fig:model_tension_future}
\end{figure*}

To further illustrate the combined constraints that could be obtained across M$_{\rm UV}$ bins from such a data set, we also re-generate Figure \ref{fig:model_tension} with 50 \textit{additional} lines of sight, and show the resulting model tensions in Figure \ref{fig:model_tension_future}. In this hypothetical scenario, the models enhancing UV-scatter, globally or with an M$_{\rm halo}$-dependence (C1, C2), as well as the global SFE model (B1) are disfavored at $>3\sigma$ significance, relative to the best-fitting sharp DMSFE model (${\rm SFE}\propto M_{\rm halo}^{0.6}$, B2). The DMSFE models are more difficult to distinguish from models boosting the UV-luminosity (A1, A2) through e.g. a more top-heavy IMF, a lack of dust attenuation, or an AGN-contribution to the UV-luminosity. Nevertheless, small differences between these models are apparent in Figure \ref{fig:model_tension_future}, and they may become statistically significant provided a much larger number of independent lines of sight.

\subsection{Impact of the Completeness Correction}
\label{sec:completeness}

Our measurements of $\sigma_{\rm CV}$ and $\mu$ rely on completeness-corrected number counts. The completeness correction is outlined in Section \ref{sec:compl_corr}, and is described in more detail in \citetalias{Weibel2026}. One worry may be that the completeness correction artificially boosts the field-to-field variance.

To directly assess the impact of the completeness correction on our inferred cosmic variance values, we repeat the bootstrapping measurement described in Section \ref{sec:bootstrapping} without applying the completeness correction. The resulting ``raw'' cosmic variance values are indeed somewhat lower than those inferred with the correction, $\sigma_{\rm CV,\,raw}=0.85_{-0.35}^{+0.33}$, $1.59_{-0.61}^{+0.59}$, and $1.95_{-0.54}^{+0.62}$ for M$_{\rm UV}<-19.5$, -20, and -20.5. However, these values are still consistent with our fiducial values within uncertainties and they do not significantly change any of our conclusions. We further note that the completeness correction significantly boosts the inferred abundance of galaxies, consistent with a large number of literature results at $z\sim10$, as shown in \citetalias{Weibel2026}.

\subsection{Reliability of Photometric Redshifts}
\label{sec:photo_zs}

Thus far we have illustrated the feasibility of measuring the galaxy bias for LBGs at $z\sim10$ from a set of spatially independent NIRCam pointings, each providing imaging in at least the F115W, F150W, F200W, F277W, F356W, and F444W filters. The success of this measurement relies on robust photometric redshifts that are available at $z\sim10$ thanks to the Lyman break shifting through the F115W filter at $z\gtrsim8.5$. The remaining 5 or more filters then all probe the rest-frame UV, making low-z interlopers unlikely, given that the measured break is strong enough to rule out Balmer break solutions at $z\sim2$, which is ensured by our stringent dropout selection (see Section \ref{sec:sample}). This is corroborated by the $>96\%$ spectroscopic confirmation rate of galaxies in the $z\sim10$ LBG sample used here.

The relatively high galaxy bias we measure, even if corrected for the impact of non-linear scales (see Section \ref{sec:comparison}) provides an additional argument for a high purity of the LBG sample from \citetalias{Weibel2026}. Uncertainties in photometric redshifts artificially weaken the clustering signal, since interloper populations are randomly distributed relative to the target population. The effect of including unassociated galaxies would dilute the measured clustering by a factor of $(1+f_{\rm interlopers})^2$ \citep[see, e.g.][]{Williams2011}. Conversely formulated, if our sample is subject to substantial contamination, this would only strengthen our tentative finding of a high cosmic variance and strong galaxy clustering at $z\sim10$.

\subsection{Combined Implications of UVLF and Clustering Measurements}
\label{sec:implications}

The findings of \citetalias{Weibel2026} suggest a rapid build-up of galaxies at cosmic dawn with $\rho_{\rm UV}$ increasing by a factor $\gtrsim50$ from $z\sim17$ to $z\sim10$. We now add the constraint that at $z\sim10$, UV-bright galaxies are highly clustered. Below, we further discuss the implications of combining these two findings in the context of theoretical models.

We first use the modeling framework from \citet{Shuntov2025b} that was already used in \citetalias{Weibel2026} to explore models that can reproduce the UVLF by either enhancing the SFE and/or $\sigma_{\rm UV}$. As shown there, a model with a slightly enhanced SFE ($\epsilon_0=0.3$, where $\epsilon_0$ is the normalization of the \textit{instantaneous} SFE, $\epsilon=\dot{M_*}/(f_b\,\dot{M}_{\rm halo})$) and $\sigma_{\rm UV}=0.6$ matches the UVLF at $z\sim10$. Here, we measure the linear galaxy bias from this model for galaxies with M$_{\rm UV}<-20$, and using Equation (14) in \citet{Shuntov2025b}, and find a bias value of 12.7, close to our measured and converted value of $14.4_{-4.8}^{+5.7}$ (see Section \ref{sec:comparison}). We also tested that enhancing $\sigma_{\rm UV}$ or $\epsilon_0$ further increases the galaxy abundance and decreases the bias, albeit to varying degrees, consistent with previous findings. In the \citet{Shuntov2025b} modeling framework, the combined measurements of the UVLF in \citetalias{Weibel2026} and the clustering in this work are consistent with a model with a somewhat enhanced SFE, and a mild contribution of star formation stochasticity. 

Returning to our models introduced in Section \ref{sec:toy_model_comparison}, we note that it is possible to accommodate some stochasticity without detecting it in the joint UVLF-clustering space. However, adding any stochasticity on top of the fiducial $M_*-{\rm M_{UV}}$ relation in \texttt{UniverseMachine} makes our fits worse. For example, a DMSFE model combined with stochasticity becomes $\sim1\sigma$ worse than a pure DMSFE model for $\sigma_{\rm UV}=0.8$, which we can therefore take as a rough upper limit on the amount of stochasticity that is consistent with our measurements in a combined DMSFE+stochasticity model. 

Relatedly, while our measurements are not constraining enough to confidently rule out any model, they tentatively disfavor models where UV-scatter is the dominant driver of the high abundance of UV-bright galaxies at $z\sim10$. In this context, \citet{Carvajal-Bohorquez2025} recently analyzed a photometric sample of galaxies at $6<z<12$ from JADES, and measured maximum values of $\sigma_{\rm UV}=0.72\pm0.02\,$mag and ${\rm SFE}=0.06\pm0.01$, concluding that neither of them is solely responsible for the high UVLF at $z>10$, consistent with our findings. Similarly \citet{Simmonds2025} compiled photometric data from JADES to characterize the star-forming main sequence and its scatter across cosmic time. They measured $\sigma_{\rm UV}=0.6-0.75$ at $6<z\leq9$, again suggesting that UV-scatter alone is unlikely to drive the observed abundance of UV-bright galaxies at $z\gtrsim9$.

Proposed physical models for the high abundance of UV bright galaxies at $z\sim10$ that do not rely on burstiness include the DMSFE \citep{Somerville2025}, which we discuss above and which provides the best fit to our measurements within our simple modeling framework. Other proposed models include the feedback-free starburst scenario \citep{Dekel2023} which implies high SFEs in the most massive halos, a lack of dust attenuation \citep{Ferrara2023}, and a more top-heavy IMF in highly star-forming regions (\citealt{Hutter2025}; see also \citealt{Mauerhofer2025}). A more in-depth analysis of the combined implications of all these models for the UVLF and the clustering strength of galaxies at $z\sim10$ is beyond the scope of this work, but our best-fit DMSFE model merits further discussion. The DMSFE scenario is motivated by cloud-scale simulations showing that SFE scales with gas surface density, and since gas surface density scales with $M_{\rm halo}$ \citep{Somerville2025}, high SFEs naturally arise in high mass halos. \citet{MBK_2025_concentration} provides a complementary theoretical underpinning, showing that gravitational acceleration within the cores of dark matter halos scales as $(1+z)^2$ at fixed halo mass, such that at $z\gtrsim8$ the interiors of sufficiently massive halos naturally exceed the critical acceleration above which gravitational collapse overcomes stellar feedback, leading to SFEs scaling with dark matter surface density and halo mass. Furthermore, it should be noted that even pre-JWST models such as the \texttt{UniverseMachine} predict a positive scaling between SFE and $M_{\rm halo}$. 
\citetalias{Weibel2026} also mentioned the model proposed by \citet{Donnan2025} who pointed out that the abundance of UV-bright galaxies at $z\gtrsim10$ may be attributed to the ever younger ages of stellar populations at high redshift, decreasing their mass-to-light ratio. Taken at face value, their model suggests a typical formation redshift of galaxies of $z\sim15$ which would then imply a rapid drop in the number density of galaxies at those redshifts, consistent with the results of \citetalias{Weibel2026}. We note here that a decrease in the mass-to-light ratio is reflected in our model implementing a global boost in the UV-luminosity (top row of panels in Figure \ref{fig:toy_models1}). This scenario is therefore consistent with our cosmic variance measurements, and is perhaps best tested through targeted searches for $z\gtrsim15$ galaxies.

\subsection{Cosmology, Assembly and Environmental Biases}
\label{sec:assembly_bias}

All of the models tested here are based on relations with one halo feature only: halo mass. However, galaxy properties and their clustering are known to be strongly associated with both internal properties beyond halo mass (e.g. age and concentration), and their exact \textit{assembly} and \textit{environments}, especially on small scales \citep{Hearin2016_assemblybias, Zentner2019_assemblybiasclustering, Jespersen2022, Chuang2024, Wu2024_largescale_env, Lim2025}. In fact, environmental bias has been shown to be the likely origin of some extreme outlier galaxies observed with JWST \citep{Jespersen2025_overdensity}. Future work should consider whether clustering amplitudes of model galaxies could be raised by e.g. preferentially populating older halos as suggested by \cite{Hearin2016_assemblybias}, while still raising the number density to the necessary levels. However, these differences can only truly be probed with constraints that include the next generation of pure parallel imaging surveys (see Figures \ref{fig:future} and \ref{fig:model_tension_future}).

It is furthermore worth noting that all models tested here assume a fixed set of cosmological parameters, which artificially shrinks the model uncertainties. Since we do not have access to \texttt{UniverseMachine} simulations with varying cosmology, we can instead estimate the impact by calculating the changes in the matter field variance while varying the fiducial WMAP cosmological parameters under 1$\sigma$ uncertainties from \cite{Planck2020}. Assuming no covariance between parameters, the resulting uncertainty on $\sigma_{\mathrm{DM}}$ is on the order of 6\%, which is subdominant to our measurement uncertainties and inter-model differences. However, this also ignores any possible feedback between cosmology and galaxy formation models.

\section{Summary and Conclusions}
\label{sec:summary_conclusions}

In this work, we use the sample of LBGs at $z\sim10$ compiled by \citetalias{Weibel2026} from a combination of legacy and pure parallel imaging from PANORAMIC to perform a measurement of their cosmic variance $\sigma_{\rm CV}$. We follow two complementary approaches to measure $\sigma_{\rm CV}$: bootstrapping the number count per field and its variance (Equation \ref{eq:cv_bootstrap}), and MCMC forward modeling of the distribution of number counts per field. This yields high values suggesting cosmic variance driven uncertainties in the galaxy number count per NIRCam pointing of $100-200$\%\ for galaxies with M$_{\rm UV}<-19.5$, consistent with values measured from the \texttt{UniverseMachine} using the same methods. Converting these measurements to a galaxy bias following \citet{Robertson2010} yields values ranging from $b_{g,\,\rm CV}\sim30-50$ for different M$_{\rm UV}$ limits. Such high bias values can be understood in the context of clustering on small, non-linear scales contributing to the observed field-to-field variance. Quantifying this effect using the \texttt{UniverseMachine}, we show that for NIRCam pointing sized fields, the galaxy bias inferred from the cosmic variance is a factor of $3-5$ higher than the linear bias for different M$_{\rm UV}$ limits.

Different models proposed to explain the high observed abundance of UV-bright galaxies at $z\gtrsim10$ differ in terms of their predictions for the clustering strength of galaxies. A direct measurement of $\sigma_{\rm UV}$ at $z\sim10$ can therefore break degeneracies between models. We explore this by implementing simple models in the \texttt{UniverseMachine} mimicking different physical processes proposed in the literature, and compare their combined predictions for the abundance and the cosmic variance of galaxies at $z\sim10$ to our measurements. Our simple models can be categorized in three classes: (1) models that boost the UV-luminosity of galaxies by $\Delta{\rm M_{UV}}$, (2) models that enhance the SFE, and (3) models that increase the scatter in the M$_{\rm UV}$-$M_h$ relation ($\sigma_{\rm UV}$). We implement two versions of each model class: (1) a global boost by $\Delta{\rm M_{UV}}$ for all galaxies, and a boost only above a certain threshold in halo mass, (2) a global constant SFE across all galaxies, and a power-law scaling of the SFE with halo mass, and (3) a global constant $\sigma_{\rm UV}$, and a scaling of $\sigma_{\rm UV}$ with halo mass (more scatter in lower mass halos, following \citealt{Sun2023}).

All these models can increase the abundance of $z\sim10$ galaxies to match our observational constraints, but simultaneously weaken the clustering signal (decrease $\sigma_{\rm UV}$) by placing galaxies at fixed M$_{\rm UV}$ in lower mass halos that are more abundant, and less clustered. Nevertheless, models differ in their combined predictions for the UVLF and the clustering. Quantifying the tension between models and the data across distinct M$_{\rm UV}$ bins (i.e. along the UVLF), we find that a global increase in the SFE, as well as models that increase $\sigma_{\rm UV}$ are disfavored at $\gtrsim2\sigma$ significance, relative to the best-fitting model that implements a scaling of the SFE with halo mass as ${\rm SFE}\propto M_{\rm halo}^{0.6}$, akin to the DMSFE proposed by \citet{Somerville2025}. With only 50 additional lines of sight, that could easily be obtained in a single dedicated pure parallel imaging survey, these model differences could be increased to $>3\sigma$.

Wider area deep NIRCam imaging along a larger number of independent lines of sight is required to obtain a more robust measurement of cosmic variance at $z\sim10$, and possibly beyond. Such data can be most efficiently acquired in the pure parallel imaging mode and yield invaluable constraints on the clustering of galaxies at cosmic dawn. They can help discriminate between different models invoked to explain the abundance of UV-bright galaxies so early in the Universe, and shed light on the physical processes driving galaxy growth in the first few hundred Myr after the Big Bang.

\section*{Acknowledgements}

We thank Oliver Hahn and Jens Stücker for the productive and informative discussion of the effect of clustering on non-linear scales on our cosmic variance measurements.
This work is based on observations made with the NASA/ESA/CSA James Webb Space Telescope. The data were obtained from the Mikulski Archive for Space Telescopes at the Space Telescope Science Institute, which is operated by the Association of Universities for Research in Astronomy, Inc., under NASA contract NAS 5-03127 for JWST. These observations are associated with program \#2514. Support for program \#2514 was provided by NASA through a grant from the Space Telescope Science Institute, which is operated by the Association of Universities for Research in Astronomy, Inc., under NASA contract NAS 5-03127. 
The Cosmic Dawn Center is funded by the Danish National Research Foundation (DNRF140). This work has received funding from the Swiss State Secretariat for Education, Research and Innovation (SERI) under contract number MB22.00072, as well as from the Swiss National Science Foundation (SNSF) through project grant 200020\_207349. The work of C.C.W. is supported by NOIRLab, which is managed by the Association of Universities for Research in Astronomy (AURA) under a cooperative agreement with the National Science Foundation. P. Dayal warmly acknowledges support from an NSERC discovery grant (RGPIN-2025-06182).

\software{
All software packages used in this work  are publicly available. We acknowledge: 
    astropy \citep{2013A&A...558A..33A,2018AJ....156..123A,2022ApJ...935..167A}, 
    matplotlib \citep{10.1109/MCSE.2007.55},  
    numpy \citep{10.1038/s41586-020-2649-2}, 
    scipy \citep{10.1038/s41592-019-0686-2}, 
    \texttt{cosmic-variance} \citep{Jespersen2025},
    \texttt{QUICKCV} \citep{Newman2014},
    the \texttt{jwst} pipeline (\citealt{10.5281/zenodo.10870758}), 
    \texttt{msaexp} (\citealt{10.5281/zenodo.7299500}), 
    and \texttt{grizli} (\citealt{10.5281/zenodo.1146904})}

\bibliography{paper}{}
\bibliographystyle{aasjournal}

\appendix

\section{Posterior Distributions of the Cosmic Variance Fits}
\label{sec:cv_posteriors}

\begin{figure*}[h!]
    \begin{center}
        \includegraphics[width=0.87\columnwidth]{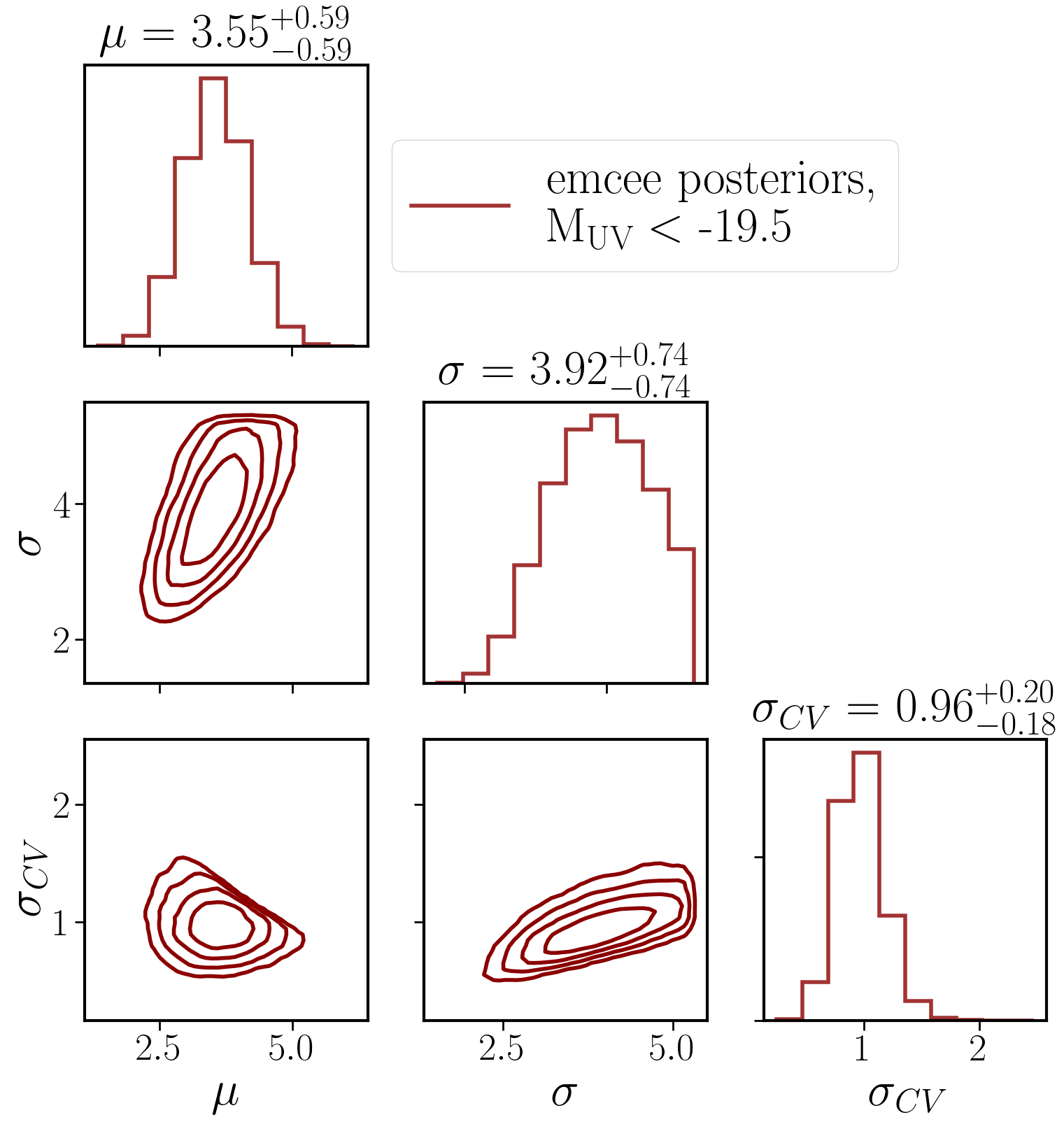}
        \centering
        \includegraphics[width=0.87\columnwidth]{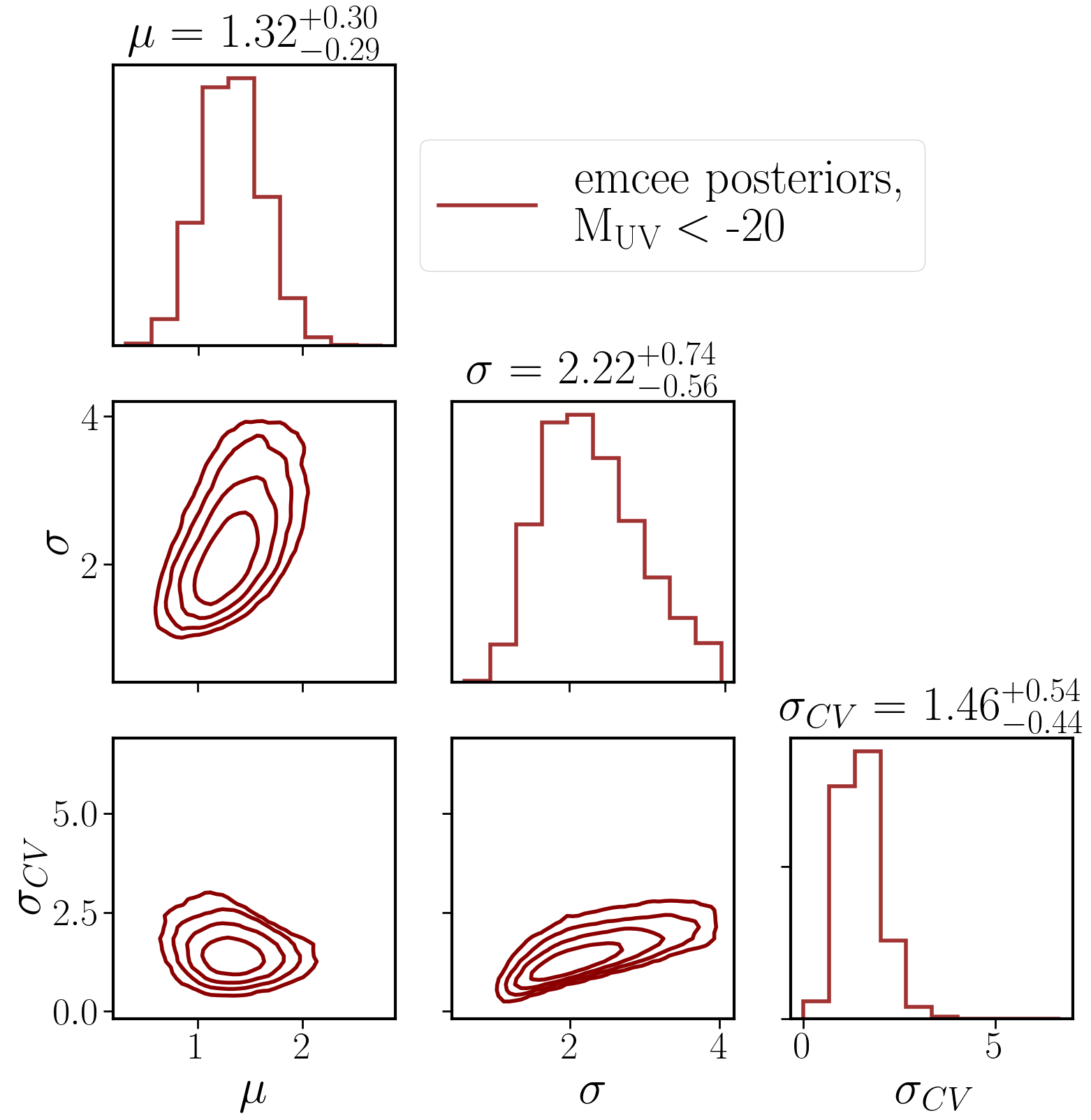}
        \centering
        \includegraphics[width=0.87\columnwidth]{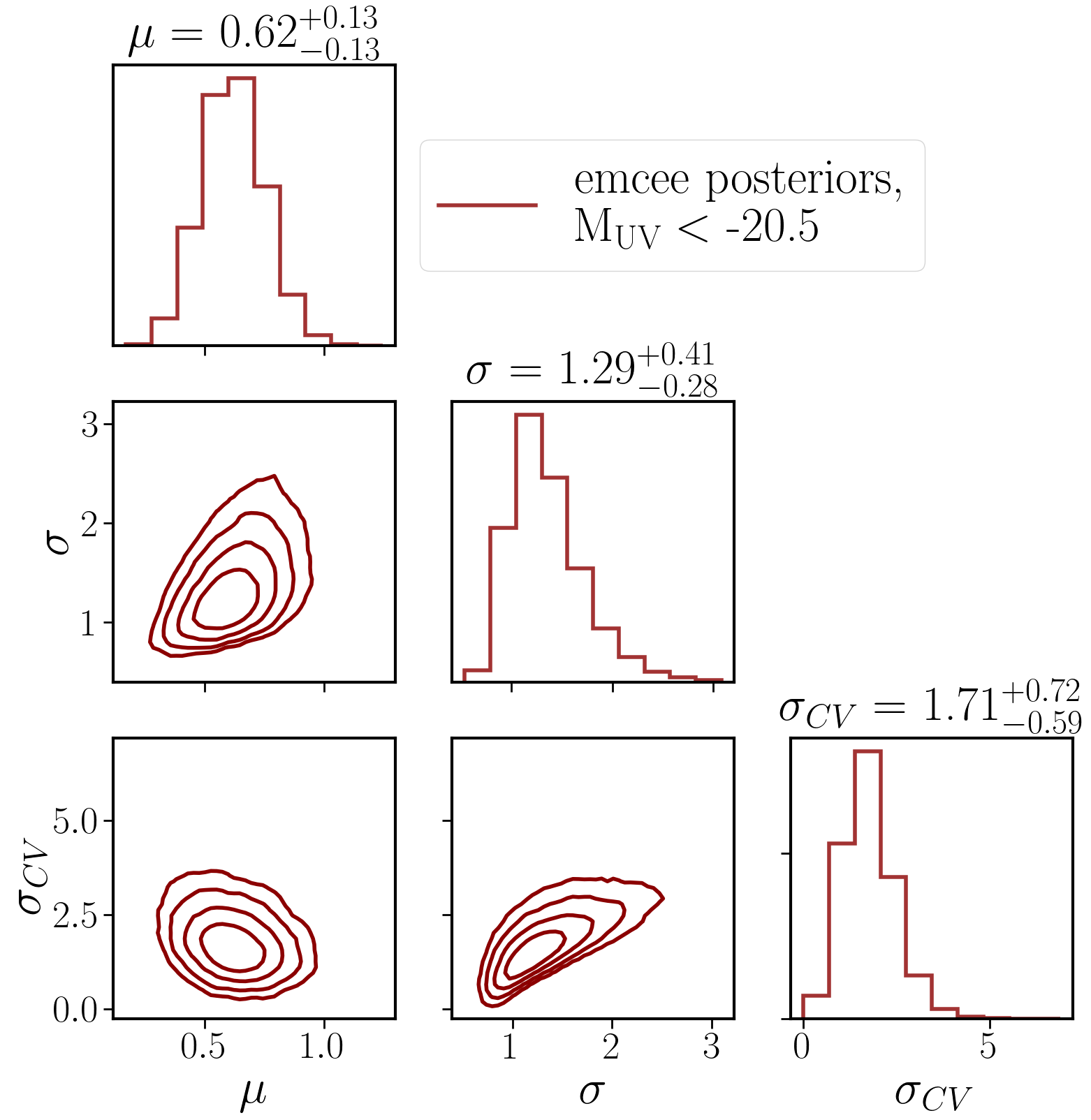}
    \end{center}
    \caption{Posterior distributions of the MCMC-fitting described in Section \ref{sec:mcmc_fitting}. The fitted quantities are the mean and variance. Contours correspond to 0.5, 1 , 1.5 and $2\sigma$ confidence regions.}
    \label{fig:cv_posteriors}
\end{figure*}

Figure \ref{fig:cv_posteriors} shows the posterior distributions of the MCMC-fits to the completeness corrected number counts of galaxies along 34 independent lines of sight, and for three different M$_{\rm UV}$ limits, as explained in Section \ref{sec:mcmc_fitting}. The widths of these distributions could potentially be underestimated due to the lack of inclusion of the uncertainty in the completeness correction. The nominal uncertainty in the completeness is small, but also likely to be underestimated \citep{Carnall2018,Jespersen2026}.

\section{Best-fitting parameters of tested models}
\label{sec:appendix_best_fit_parameters}

In Table \ref{tab:model_params}, we list the constraints we get on the model parameters for the different tested models. Note that the quoted uncertainties are calculated by considering the typical 0.5 value increase in the log-likelihood on either side, which \textit{does not} take into account any model misspecification corrections for a model that does not actually provide a good fit. As discussed in \cite{Jespersen2026}, this results in the quoted uncertainties being underestimated, and the central values being biased. The more misspecified the model, the more severe the issue, which is why we have focused on the task of determining \textit{which} model is most appropriate, rather than constraining the parameters that characterize a specific model.

\begin{table}[h!]
    \centering
    \setlength\tabcolsep{0.1cm}
    \begin{tabular}{lll}
        Model Name & Parameter & Best-fit value \\
        & & \\[\dimexpr-\normalbaselineskip+1pt]
        \hline
        \hline
        & & \\[\dimexpr-\normalbaselineskip+1pt]
        Global $\Delta$M$_{\rm UV}$ (A1) & $\Delta\mathrm{M_{UV}}$ & $1.02^{+0.08}_{-0.07}$ \\
        & & \\[\dimexpr-\normalbaselineskip+1pt]
        \hline
        & & \\[\dimexpr-\normalbaselineskip+1pt]
        $\Delta$M$_{\rm UV}$: $M_{\mathrm{halo}} \geq 10^{10} M_{\odot}$ (A2) & $\Delta\mathrm{M_{UV}}$ & $1.01^{+0.08}_{-0.08}$ \\
        & & \\[\dimexpr-\normalbaselineskip+1pt]
        \hline
        & & \\[\dimexpr-\normalbaselineskip+1pt]
        $\Delta$M$_{\rm UV}$: $M_{\mathrm{halo}} \geq 10^{10.5} M_{\odot}$ (A2) & $\Delta\mathrm{M_{UV}}$ & $0.93^{+0.13}_{-0.19}$ \\
        & & \\[\dimexpr-\normalbaselineskip+1pt]
        \hline
        & & \\[\dimexpr-\normalbaselineskip+1pt]
        Global SFE (B1) & SFE & $0.098^{+0.004}_{-0.006}$ \\
        & & \\[\dimexpr-\normalbaselineskip+1pt]
        \hline
        & & \\[\dimexpr-\normalbaselineskip+1pt]
        DMSFE: ${\rm SFE}\propto M_{\rm halo}^{0.5}$ (B2) & $\mathrm{SFE_{peak}}$ & $0.63^{+0.03}_{-0.02}$ \\
        & & \\[\dimexpr-\normalbaselineskip+1pt]
        \hline
        & & \\[\dimexpr-\normalbaselineskip+1pt]
        DMSFE: ${\rm SFE}\propto M_{\rm halo}^{0.6}$ (B2) & $\mathrm{SFE_{peak}}$ & $0.94^{+0.06}_{-0.06}$ \\
        & & \\[\dimexpr-\normalbaselineskip+1pt]
        \hline
        & & \\[\dimexpr-\normalbaselineskip+1pt]
        Global $\sigma_{\mathrm{UV}}$ (C1) & $\sigma_\mathrm{UV}$ & $1.21^{+0.04}_{-0.04}$ \\
        & & \\[\dimexpr-\normalbaselineskip+1pt]
        \hline
        & & \\[\dimexpr-\normalbaselineskip+1pt]
        $M_{\rm halo}$-dependent $\sigma_{\rm UV}$ (C2) & $\sigma_\mathrm{UV,\,norm}$ & $1.11^{+0.06}_{-0.05}$ \\
        & & \\[\dimexpr-\normalbaselineskip+1pt]
        \hline
    \end{tabular}
    \caption{The best-fittting parameters for all tested models.}
    \label{tab:model_params}
\end{table}

\section{Model Comparison in $\mathbf{\mathrm{M_{UV}}}$ Bins}
\label{sec:appendix_model_comp_bins}

\begin{figure*}
    \centering
        \includegraphics[width=\textwidth]{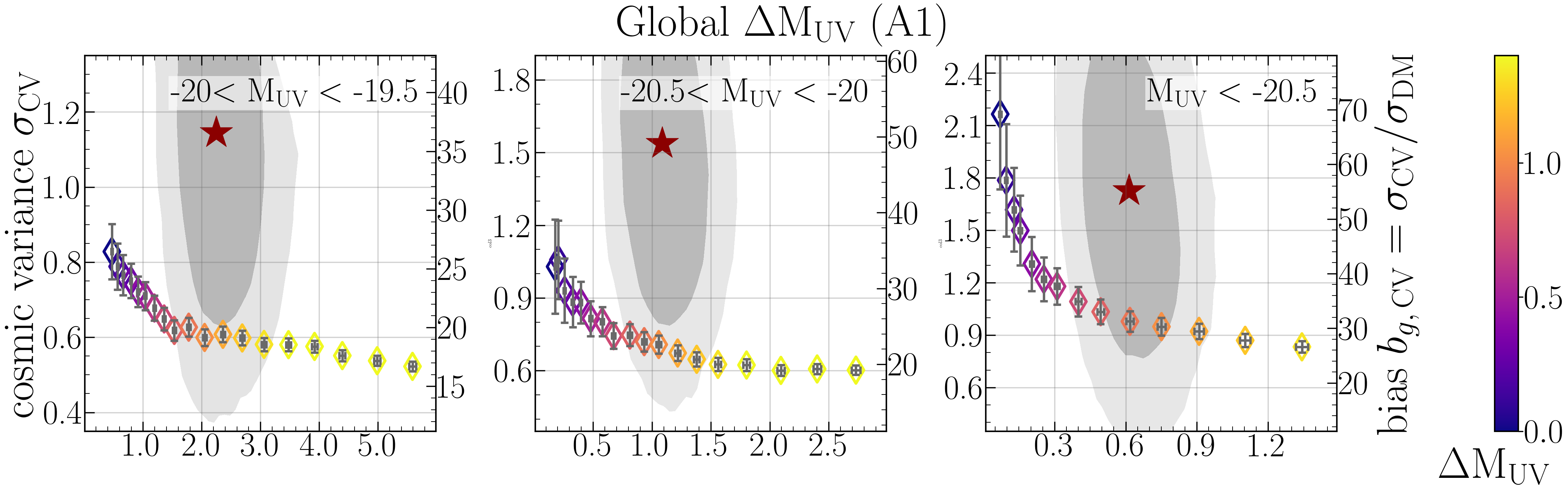}
        \includegraphics[width=\textwidth]{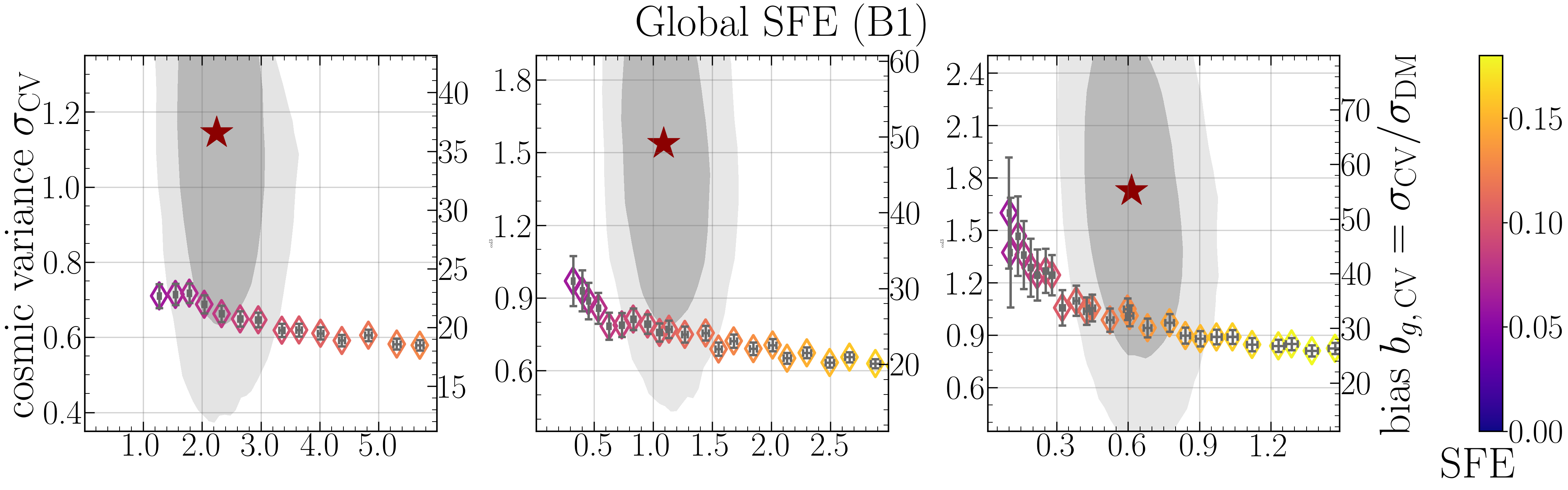}
        \includegraphics[width=\textwidth]{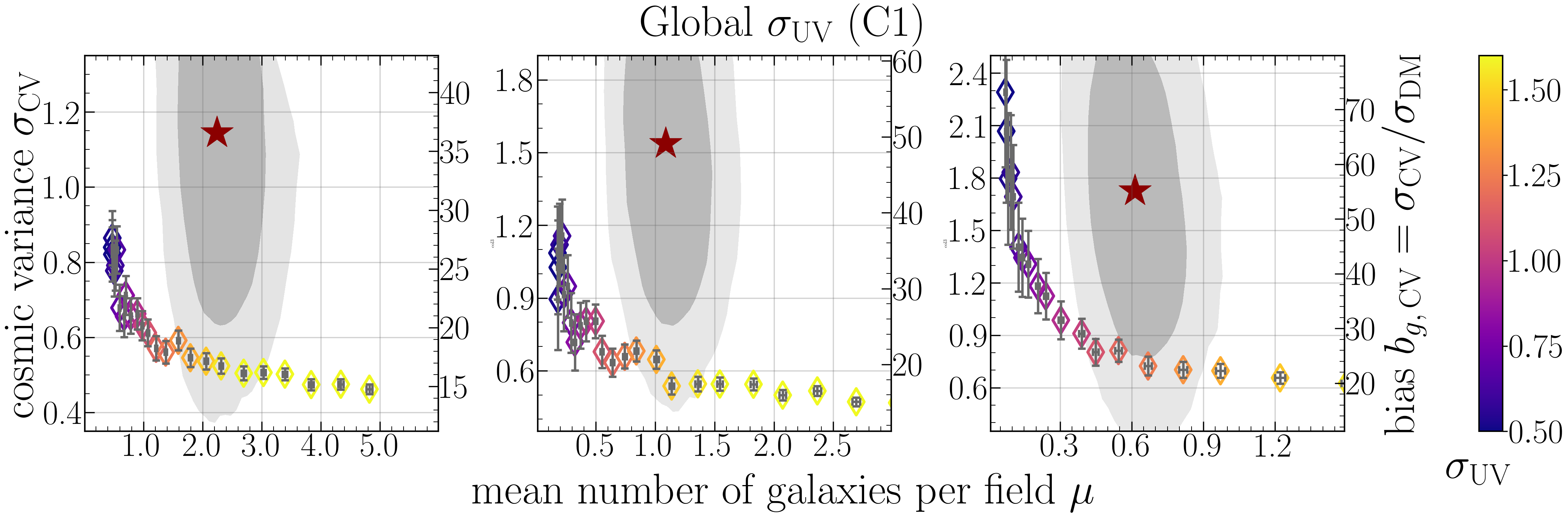}
    \caption{Same as Figure \ref{fig:toy_models1}, but in bins of M$_{\rm UV}$, instead of for different M$_{\rm UV}$ limits, as indicated in the top row of panels.}
    \label{fig:toy_models1_bins}
\end{figure*}

\begin{figure*}
    \centering
        \includegraphics[width=\textwidth]{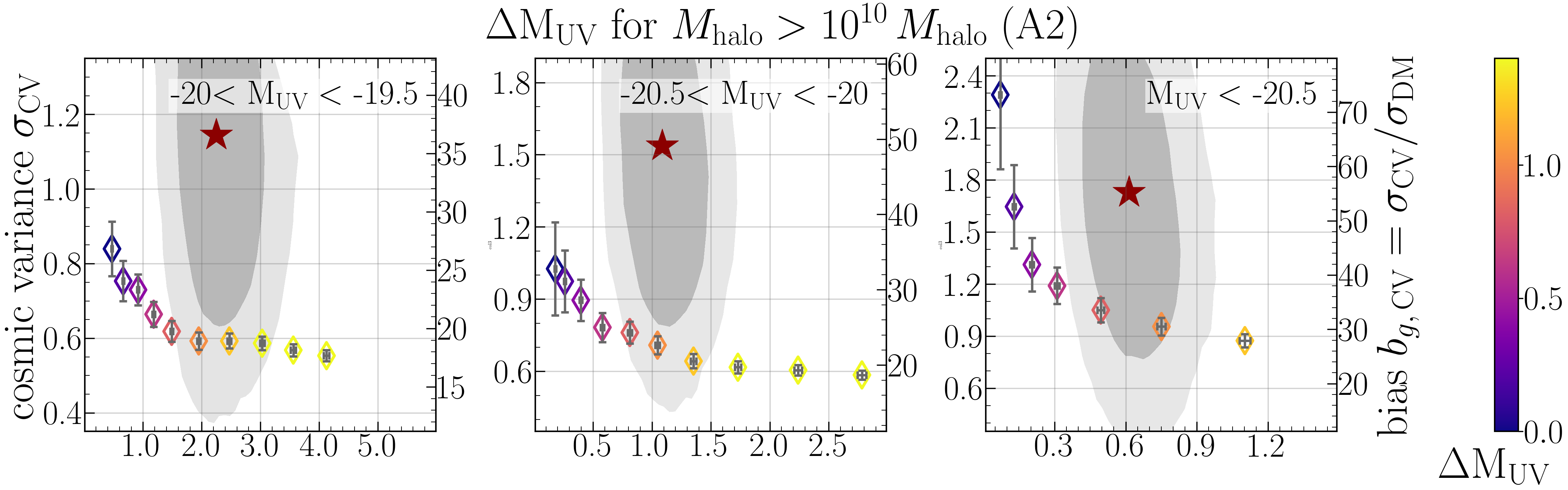}
        \includegraphics[width=\textwidth]{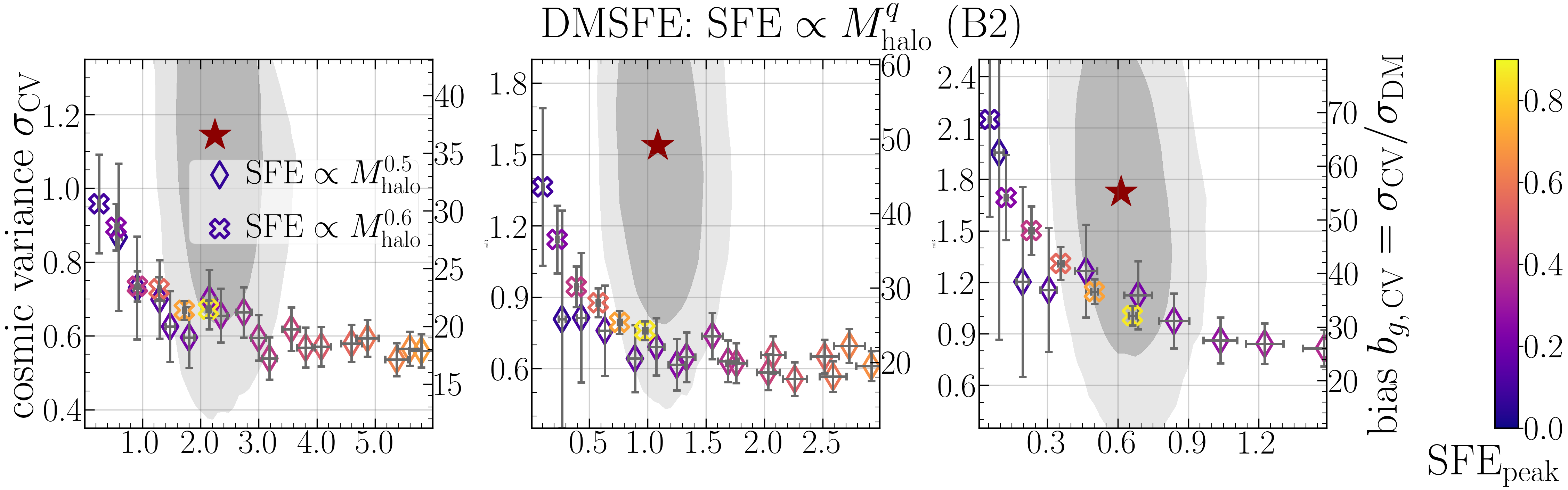}
        \includegraphics[width=\textwidth]{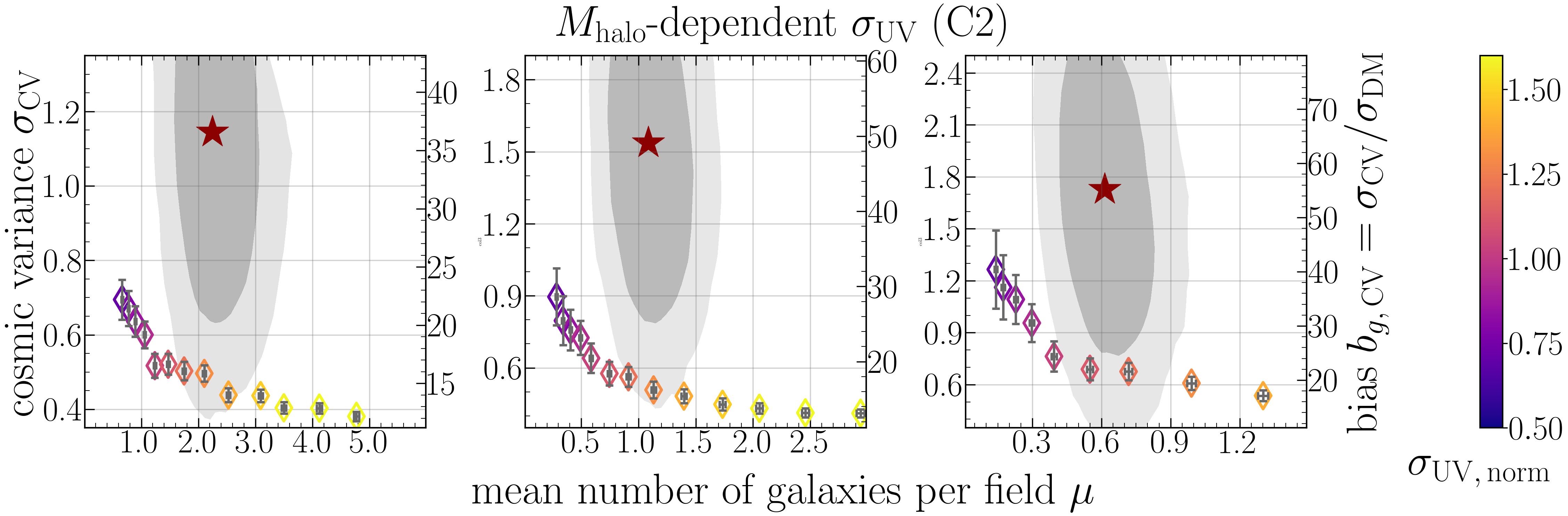}
    \caption{Same as Figure \ref{fig:toy_models2}, but in bins of M$_{\rm UV}$, instead of for different M$_{\rm UV}$ limits, as indicated in the top row of panels. We do not show the model boosting M$_{\rm UV}$ for $M_{\rm halo}>10^{10.5}\,{\rm M_\odot}$ (A2) because this has no effect on galaxies in the faintest bin that reside in lower mass halos.}
    \label{fig:toy_models2_bins}
\end{figure*}

In Figures \ref{fig:toy_models1_bins} and \ref{fig:toy_models2_bins}, we show our combined cosmic variance and number density measurements, as well as the model trajectories for the models introduced in Section \ref{sec:toy_model_comparison}, but in bins of M$_{\rm UV}$, instead of for different M$_{\rm UV}$ limits. The resulting measurements in the fainter bins are less constraining compared to the corresponding limits due to the smaller number of sources (and sufficiently deep fields). The brightest bin corresponds to M$_{\rm UV}<-20.5$, and is therefore identical to the brightest limit used throughout this paper. We use the difference between our measurements and the different models in bins to compute the combined model tension in Section \ref{sec:model_tension}, so as to avoid galaxies contributing to more than one bin. We do not show the model boosting the UV-luminosity in halos with $M_{\rm halo}>10^{10.5}\,{\rm M_\odot}$ (A2) because this has almost no effect on the number density of galaxies in the faintest bin.

\end{document}